\documentclass{ieeetmlcn}
\usepackage{cite}
\usepackage{dblfloatfix}

\usepackage{subcaption}
\usepackage{overpic}
\usepackage{amsmath,amssymb,amsfonts,amsthm,bm}
\usepackage{graphicx}
\usepackage{epstopdf}
\usepackage{textcomp}
\usepackage{soul,xcolor}
\usepackage{float}
\usepackage{multirow}
\usepackage{multicol}
\usepackage{algorithm}
\usepackage{algpseudocode}
\algnewcommand{\LongState}[1]{%
  \State
  \parbox[t]{\dimexpr\linewidth-\algorithmicindent-1.5em\relax}{%
    \raggedright #1}}

\usepackage{comment}
\algrenewcommand\algorithmicrequire{\textbf{Input:}}
\algrenewcommand\algorithmicensure{\textbf{Output:}}

\theoremstyle{plain}

\newtheorem{remark}{Remark}

\def\Htran{\mbox{\tiny $\mathrm{H}$}}
\def\Ttran{\mbox{\tiny $\mathrm{T}$}}

\usepackage{tikz}
\usepackage{hyperref}
\hypersetup{hidelinks}
\usetikzlibrary{arrows.meta,positioning,calc,backgrounds}
\definecolor{ulc}{RGB}{0,102,170}    
\definecolor{dlc}{RGB}{190,60,30}    
\definecolor{gnnc}{RGB}{90,60,150}   
\definecolor{offc}{RGB}{160,160,160} 

\def\BibTeX{{\rm B\kern-.05em{\sc i\kern-.025em b}\kern-.08em
    T\kern-.1667em\lower.7ex\hbox{E}\kern-.125emX}}
\AtBeginDocument{\definecolor{tmlcncolor}{cmyk}{0.93,0.59,0.15,0.02}%
  \definecolor{NavyBlue}{RGB}{0,86,125}}
\def\OJlogo{\vspace{-4pt}\includegraphics[height=18pt]{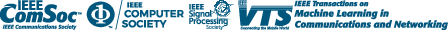}}
\def\seclogo{\vspace{10pt}\includegraphics[height=18pt]{new_logo}}
\def\authorrefmark#1{\ensuremath{^{\textbf{#1}}}}
\begin{document}
\markboth{}{Ramezani and Demir: GNN-Aided Joint Uplink--Downlink Fronthaul Bit Allocation}
\title{GNN-Aided Joint Uplink--Downlink Fronthaul Bit Allocation for Cell-Free Massive MIMO}
\author{Parisa Ramezani\authorrefmark{1} and \"Ozlem Tu\u{g}fe Demir\authorrefmark{2}}
\affil{School of Electrical, Mechanical and Biomedical Engineering, University of Technology Sydney, Sydney, Australia}
\affil{Department of Electrical and Electronics Engineering, Bilkent University, Ankara, T\"urkiye}
\corresp{Corresponding author: Parisa Ramezani (email: parisa.ramezani@uts.edu.au).}
\authornote{Claude Opus 5.5 (Anthropic) was used to assist with the simulation code and the drafting of the manuscript text. The authors reviewed and edited all content and take full responsibility for it. The work by \"O. T. Demir was supported by the Scientific and Technological Research Council of T\"urkiye (TUBITAK) BIDEB 2232-B International Fellowship for Early Stage Researchers under Grant Number 122C149. The work by \"O. T. Demir was supported by the BAGEP Award of the Science Academy with funding supplied by \"Ozg\"ul ve Arma\u{g}an \c{C}a\u{g}layan Vakf\i.}
\begin{abstract}
The finite capacity of fronthaul links is an important practical limitation in
cell-free massive multiple-input multiple-output (MIMO) systems, where the
distributed access points (APs) must exchange information with a central
processing unit (CPU). In this paper, we investigate joint uplink--downlink
fronthaul bit-allocation in a centralized cell-free massive MIMO system. The
uplink fronthaul bits determine the quantization quality of the pilot
observations forwarded to the CPU and thereby affect channel estimation and
downlink precoding, whereas the downlink bits determine the quantization
distortion of the precoding vectors conveyed to the APs, which form their transmit signals locally from the data symbols. We formulate the joint
allocation of these quantization resolutions under a shared fronthaul capacity
constraint with the objective of maximizing the expected downlink sum rate.
To address the resulting combinatorial integer optimization problem, we first
develop a greedy bit-allocation algorithm that generates locally optimized
solutions. These solutions are then used to train a permutation-equivariant
bipartite graph neural network (GNN) that learns the mapping from large-scale
fading coefficients to feasible fronthaul bit-allocations. The additive quantization noise model is employed during
optimization and training, while the actual Lloyd--Max quantizers are used for the spectral efficiency evaluation. Finally, we
evaluate the effectiveness of the proposed GNN-based approach in terms of sum
spectral efficiency, generalization capability, and runtime.
\end{abstract}
\maketitle
\section{Introduction}
Cell-free massive multiple-input multiple-output (MIMO) has emerged as a
promising distributed network architecture for providing uniform high-quality
wireless service over a large coverage area. In contrast to conventional
cellular networks, where each user equipment (UE) is primarily associated with
a particular base station or cell, cell-free massive MIMO relies on a large
number of geographically distributed access points (APs) that cooperatively
serve the UEs over the same time--frequency resources \cite{Ngo2017b, Nayebi2017a,Interdonato2018,ozdogan2019performance,Bjornson2020a}. The APs are coordinated
through one or more central processing units (CPUs). The absence of
fixed cell boundaries enables the network to exploit signals received or
transmitted from multiple spatially separated APs. 

The distributed nature of cell-free massive MIMO provides substantial
macro-diversity, as it increases the chance for a UE to have favorable propagation conditions to some of the available APs. Moreover, the large number of
distributed antennas can provide favorable propagation and coherent
beamforming gains, while mitigating the strong performance variations and
inter-cell interference associated with conventional cell-centric networks.
As a result, cell-free massive MIMO can provide more uniform coverage and
substantial improvements in the service experienced by poorly-served UEs,
such as those located near conventional cell edges. These benefits, however,
come at the expense of increased coordination and signal exchange between the
distributed APs and the CPUs, making the capacity and efficient utilization of
the fronthaul network important considerations in practical cell-free massive
MIMO deployments.

Although ideal or unlimited fronthaul capacity is commonly assumed in the
cell-free massive MIMO literature for analytical tractability, practical
fronthaul links have finite capacity. Therefore, the available fronthaul
resources must be carefully utilized and allocated to avoid performance
degradation and to fully exploit the gains of distributed cooperation.

In this paper, we account for the finite
fronthaul capacity by jointly allocating the available fronthaul quantization
bits between uplink pilot transmission and downlink data transmission under a
shared fronthaul budget. Instead of forwarding the precoded baseband samples, whose fronthaul load grows with the number of downlink channel uses in each coherence block, the CPU conveys the quantized precoding vectors once per coherence block together with the data symbols, and each AP forms its transmit signal locally. Since the number of UEs is typically much smaller than the number of downlink channel uses per coherence block, this functional split considerably reduces the downlink fronthaul load.
The allocated uplink fronthaul bits determine the quantization quality of the pilot observations and the accuracy of the channel estimates obtained at the CPU. The downlink bits control the quantization distortion of the precoding vectors
sent from the CPU to the APs over the capacity-limited fronthaul links.
Hence, the uplink and downlink fronthaul allocations affect the downlink
performance in different but coupled ways: the uplink allocation influences
the quality of the channel state information (CSI) used for precoding, whereas the
downlink allocation determines the accuracy of the precoding vectors with which the APs form their transmit signals. This coupling motivates a joint
allocation of the available fronthaul resources. 
\subsection{Related Work}
Several works in the literature have investigated wireless systems with
capacity-constrained fronthaul/backhaul links and studied different strategies
for mitigating the resulting quantization or compression distortions through
resource allocation, signal compression, and quantization-aware transceiver
design.

In \cite{Park2013joint}, the downlink of a cloud radio access network with
finite-capacity backhaul links was considered, where the precoded baseband
signals were compressed before being conveyed to the base stations.
The precoding matrix and the covariance of the compression noise were jointly
optimized for weighted sum-rate maximization under backhaul-capacity and
transmit-power constraints.
Reference \cite{Zhou2014optimized} studied the uplink of a cloud radio access network with finite-capacity backhaul links, where the base stations compress their received signals before forwarding them to the central processor. The backhaul limitation was incorporated through rate-distortion constraints, and the quantization noise levels were optimized for weighted sum-rate maximization under a sum backhaul-capacity constraint. Both Wyner--Ziv and independent single-user compression strategies were considered, and low-complexity optimization methods were developed for the corresponding compression design.

The impact of capacity-limited fronthaul has also been investigated in cell-free massive MIMO systems, particularly in connection with the compression of uplink or downlink signals exchanged between the APs and the CPU. In \cite{Bashar2019b}, the uplink max--min rate was optimized under fronthaul-capacity constraints by considering quantized channel estimates, received signals, or locally processed signals at the APs, with the quantization effect modeled using the Bussgang decomposition. The authors jointly designed the receiver filtering and UE power allocation and further proposed a fronthaul-aware UE assignment strategy. 
A different uplink study in \cite{Masoumi2020a} examined several AP processing architectures, including compress-forward-estimate, estimate-compress-forward, and estimate-multiply-compress-forward. The available fronthaul capacity was shared between the compressed CSI and data signals, and fronthaul-rate allocation and UE power control were optimized to improve the achievable sum rate.
For the downlink, \cite{Boroujerdi2019a} considered a cell-free massive MIMO system in which the centrally precoded signals are compressed before transmission to the distributed antennas over finite-capacity fronthaul links. The authors formulated a joint power-weight allocation and compression problem and also investigated separate power allocation and compression as a lower-complexity alternative.
In \cite{Topal2024energy}, an energy-efficient cell-free massive MIMO system with
wireless fronthaul was considered, where a joint antenna-activation and power-allocation
scheme was developed to minimize the end-to-end network power consumption
under UE quality-of-service and fronthaul-capacity constraints. 
More recently, a joint quantized precoding and bit-allocation framework for
fronthaul-constrained cell-free massive MIMO was proposed in \cite{demir2026joint}. The
centrally designed downlink precoder is quantized before being conveyed to the
APs, with the quantization distortion modeled using Bussgang
decomposition. The precoder and the per-AP per-UE quantization resolutions are
jointly optimized under a total fronthaul bit budget and per-AP power
constraints using a block coordinate descent algorithm.

Related works have also investigated fronthaul limitations in conventional
and reconfigurable intelligent surface (RIS)-assisted multi-UE (MU)-MIMO systems, where the finite-capacity fronthaul requires
the centrally computed precoding coefficients to be represented with finite
resolution before being conveyed to the antenna system.
A quantization-aware precoding framework was developed in \cite{khorsandmanesh2023optimized} for
downlink MU-MIMO with capacity-constrained fronthaul. Rather than first
designing a continuous precoder and subsequently quantizing it, the precoding
matrix is directly optimized over the finite fronthaul quantization alphabet
to minimize the receive-side mean square error (MSE). Sphere-decoding-based and lower-complexity
heuristic methods were proposed to solve the resulting discrete precoding
problem.
The impact of fronthaul quantization was subsequently considered together with
finite-resolution RIS configuration in RIS-assisted MU-MIMO systems. In
\cite{Ramezani2024MSE}, the sum MSE is minimized by jointly designing the quantized
precoding vectors, discrete RIS phase shifts, and receiver gains. A block
coordinate descent framework is employed, with the discrete precoding and RIS
configuration subproblems solved using sphere decoding.
Extending this line of work to sum-rate maximization, \cite{Ramezani2025joint} jointly
optimizes the fronthaul-quantized precoder and the RIS configuration in a
downlink RIS-assisted MU-MIMO system by reformulating the resulting problem through the weighted-MSE equivalence. 
Recently, a splitting-precoding architecture was proposed in \cite{Khorsandmanesh2026splitting}
for massive MIMO systems with limited fronthaul capacity. Instead of
transmitting a full-dimensional precoding matrix over the fronthaul, the radio unit
first performs local subspace selection, while the baseband unit computes a
quantization-aware refinement precoder in the resulting lower-dimensional
subspace. This reduces the fronthaul load and allows a higher quantization
resolution for a given fronthaul budget.

Although considerable effort has been devoted to accounting for limited
fronthaul capacity across different network architectures, transmission
directions, and compression or quantization models, the joint allocation of
fronthaul resources between the uplink and downlink has yet to be thoroughly investigated. This is important because optimizing the uplink and downlink allocations separately may lead to an inefficient use of the available fronthaul capacity when both directions must share a common fronthaul budget. In our recent work \cite{demir2026GNN}, we took some initial steps toward joint
uplink--downlink fronthaul design by optimizing the bit-allocation together
with split combining and split precoding under a common fronthaul budget. However, that work
assumed perfect CSI and did not account for the impact of uplink fronthaul
quantization on channel estimation and, consequently, on downlink precoding.
This leaves a fundamental gap since the uplink
fronthaul resources allocated to pilot observations directly affect the
quality of the channel estimates and the effectiveness of downlink precoding. 
This paper addresses this gap by jointly optimizing the uplink and downlink
fronthaul bit-allocation under a shared fronthaul capacity constraint, accounting for the impact of
uplink quantization on channel estimation and that of downlink quantization on the
precoding vectors. To handle the resulting combinatorial problem, we utilize a permutation-equivariant bipartite graph neural network (GNN) as it naturally captures the UE--AP interaction structure while being equivariant to any relabeling of the UEs and the APs.

Over the past decade, neural networks have been increasingly explored for
wireless resource optimization as a means of replacing computationally
intensive real-time optimization procedures with fast learned mappings. For
example, in \cite{Lee2018resource}, a deep neural network (DNN)-based resource allocation scheme was
developed for multi-channel underlay cognitive radio networks, where channel
gain information is directly mapped to the transmit power allocation of the
secondary UEs. The DNN was trained to improve the spectral efficiency of the
secondary UEs while limiting the interference caused to the primary UEs. 
In \cite{Sun2018learning}, a DNN was trained to approximate the
input--output relationship of the weighted minimum mean squared error algorithm, mapping channel
realizations to power allocations, and hence substantially reducing the online computational burden.
Among the various neural network architectures, GNNs have attracted
particular attention for wireless resource allocation because they can
explicitly exploit the underlying network topology while preserving permutation
equivariance with respect to the ordering of network entities. In \cite{eisen2020optimal}, the authors proposed a random edge GNN  for wireless resource allocation,
where the fading interference pattern is represented as a graph and the
resource allocation policy is learned through an unsupervised, model-free approach. The resulting architecture is scalable and can be
transferred across networks with different sizes and topologies. Reference
\cite{shen2021gnn} modeled wireless networks as channel graphs and
developed message-passing GNNs for radio resource management problems,
including power control and beamforming. The framework exploits generalization to larger network sizes, while also
establishing a theoretical connection between the proposed GNNs and
distributed optimization algorithms. The authors in \cite{Lee2021graph} proposed a
graph-embedding-based approach for link scheduling in device-to-device (D2D) networks, where
D2D pairs are represented as graph nodes and interference links as edges.
The resulting node embeddings are used to determine link activation, allowing 
the scheduling decisions to be learned with  few training network
layouts and without requiring accurate CSI.

\subsection{Contributions}
The main contributions of this paper are summarized as follows:
\begin{itemize}

\item We develop a joint uplink--downlink fronthaul bit-allocation framework
for centralized cell-free massive MIMO systems. Instead of treating the two
transmission directions independently, the proposed formulation accounts for
the impact of uplink pilot quantization on channel estimation and downlink precoding, as well as the direct impact of downlink quantization on
the precoding vectors conveyed to the APs. The uplink and downlink quantization resolutions are
jointly optimized under a shared fronthaul capacity constraint to maximize the
expected downlink sum rate.

\item To address the resulting combinatorial integer optimization problem, we
first develop a greedy bit-allocation algorithm that accounts for the different
fronthaul costs of uplink and downlink quantization and generates locally
optimized allocation solutions. These solutions are then used to train a
permutation-equivariant bipartite GNN that learns the mapping from the
large-scale fading coefficients to the fronthaul bit-allocation. The proposed
GNN exploits the UE--AP interaction structure and enables fast online
bit-allocation without repeatedly solving the combinatorial problem.

\item Numerical results demonstrate that the proposed GNN-based allocation
substantially outperforms baseline
strategies while closely following the performance of the greedy
solution. Moreover, the trained GNN generalizes to different numbers of APs
and UEs without retraining and achieves a substantially lower online runtime
than the greedy algorithm, demonstrating its suitability for real-time
fronthaul resource allocation.
\end{itemize}

\subsection{Organization}
The remainder of this paper is organized as follows. Section~\ref{section:sysmod}
introduces the system model and formulates the joint uplink--downlink fronthaul
bit-allocation problem. Building on this formulation, Section~\ref{section:greedy}
develops the greedy bit-allocation algorithm used to generate the training
labels. The proposed permutation-equivariant bipartite GNN and its learning
framework are then presented in Section~\ref{section:proposed}. To characterize
the quantization effects more accurately, Section~\ref{section:Bussgang}
derives the achievable spectral efficiency with the actual Lloyd--Max quantizers in the uplink and downlink. Numerical results
evaluating the proposed approach, including its performance, generalization,
and computational efficiency, are reported in Section~\ref{section:results}.
Finally, Section~\ref{section:conclusion} concludes the paper.

\begin{figure*}
\centering
\begin{tikzpicture}[
  font=\footnotesize, >={Latex[length=2mm]},
  blk/.style={draw, rounded corners=2pt, fill=white, align=center,
              minimum height=8mm, inner sep=3pt},
  qnt/.style={draw, fill=white, minimum width=7mm, minimum height=4.5mm,
              inner sep=0pt, font=\scriptsize},
  ap/.style={draw, thick, rounded corners=2pt, fill=gray!8, align=center,
             minimum width=19mm, minimum height=8mm, inner sep=2pt},
  ul/.style={->, thick, ulc},
  dl/.style={->, thick, dlc},
  ch/.style={densely dashed, gray!70},
]

\tikzset{
  antenna/.pic={
    \draw[thick] (0,0) -- (0,-0.32);
    \draw[thick, fill=white] (0,-0.32) -- (-0.11,-0.50) -- (0.11,-0.50) -- cycle;
  },
  ue/.pic={
    \draw[thick, rounded corners=1.2pt, fill=white] (-0.15,-0.27) rectangle (0.15,0.27);
    \draw (-0.10,-0.12) rectangle (0.10,0.20);
    \fill (0,-0.20) circle (0.02);
  },
}

\draw[very thick, rounded corners=4pt, fill=gray!4] (0,4.2) rectangle (13.6,8.0);
\node[anchor=north west, font=\small\bfseries] at (0.1,7.95) {CPU};

\node[blk, text width=28mm] (est) at (2.4,6.9)
  {AP-wise LMMSE\\channel estimation};
\node[blk, text width=26mm] (prec) at (6.6,6.9)
  {Centralized\\precoding $\{\mathbf{w}_k\}$};
\node[blk, text width=30mm, draw=gnnc, fill=gnnc!8, very thick] (gnn) at (11.3,7.15)
  {Bipartite GNN\\[-1pt] {\scriptsize $\{\beta_{kl}\}\mapsto(T_l,q_l,p_l)$}};
\node[blk, text width=34mm, draw=gnnc, minimum height=7mm] (proj) at (11.3,5.95)
  {Integer projection:\\[-1pt]
   {\scriptsize $\frac{1}{\tau_c}\sum_{l}\big(\tau_p b_l^{\mathrm{UL}}+K b_l^{\mathrm{DL}}\big)\le B_{\mathrm{tot}}$}};

\draw[->, thick] (est) -- node[above]{$\{\hat{\mathbf{h}}_k\}$} (prec);
\draw[->, thick, gnnc] (gnn) -- (proj);
\draw[<-, thick, gnnc] (gnn.north) -- ++(0,0.85)
  node[above, align=center, text=black]{large-scale fading $\{\beta_{kl}\}$};
\draw[->, thick, gnnc, densely dashed] (proj.west) -- ++(-0.55,0)
  node[left, text=gnnc, font=\scriptsize]{$\mathbf{b}\rightarrow\{{\mathcal{Q}_l^{\mathrm{UL}}},{\mathcal{Q}_l^{\mathrm{DL}}}\}$};

\node[ulc] (yin) at (2.4,5.75) {$\{\tilde{\mathbf{y}}_{t,l}\}$};
\draw[ul] (yin) -- (est);

\def\xA{2.0}  \def\xB{6.6}  \def\xC{11.3}

\draw[thick, dlc] (prec.south) -- (6.6,5.35);
\draw[thick, dlc] ($(\xA,5.35)+(0.45,0)$) -- ($(\xC,5.35)+(0.45,0)$);
\node[dlc, anchor=south] at (4.3,5.35) {$\{\mathbf{W}_l\},\{s_k\}$};

\foreach \x/\l/\c/\st in {\xA/1/black/solid, \xB/l/black/solid, \xC/L/offc/densely dashed}{
  \node[qnt, draw=\c] (qd\l) at ($(\x,4.75)+(0.45,0)$) {{$\mathcal{Q}_{\l}^{\mathrm{DL}}$}};
  \draw[thick, dlc, \st] ($(\x,5.35)+(0.45,0)$) -- (qd\l.north);
  \node[anchor=west, font=\scriptsize, text=\c] at (qd\l.east) {$b_{\l}^{\mathrm{DL}}$};
  \node[ap, draw=\c, \st] (ap\l) at (\x,1.95) {AP $\l$\\[-2pt]{\scriptsize pilot despreading}\\[-2pt]{\scriptsize apply precoding}};
  \foreach \a in {-0.6,-0.2,0.2,0.6}{ \pic[\c] at ($(ap\l.south)+(\a,0)$) {antenna}; }
  \node[qnt, draw=\c] (qu\l) at ($(\x,2.95)-(0.45,0)$) {{$\mathcal{Q}_{\l}^{\mathrm{UL}}$}};
  \node[anchor=east, font=\scriptsize, text=\c] at (qu\l.west) {$b_{\l}^{\mathrm{UL}}$};
}

\foreach \x/\l in {\xA/1, \xB/l}{
  \draw[thick, ulc] ($(ap\l.north)-(0.45,0)$) -- (qu\l.south);
  \draw[ul] (qu\l.north) -- ($(\x,4.2)-(0.45,0)$)
     node[pos=0.5, left, font=\scriptsize]{$\tilde{\mathbf{y}}_{t,\l}$};
  \draw[dl] (qd\l.south) -- ($(ap\l.north)+(0.45,0)$)
     node[pos=0.45, right, font=\scriptsize]{$\widetilde{\mathbf{W}}_{\l}$};
}
\draw[thick, offc, densely dashed] ($(apL.north)-(0.45,0)$) -- (quL.south);
\draw[thick, offc, densely dashed] (quL.north) -- ($(\xC,4.2)-(0.45,0)$);
\draw[thick, offc, densely dashed] (qdL.south) -- ($(apL.north)+(0.45,0)$);
\node[offc, font=\scriptsize, align=left, anchor=west] at ($(apL.east)+(0.05,0.1)$)
  {inactive:\\$b_L^{\mathrm{UL}}\!=b_L^{\mathrm{DL}}\!=0$};

\node at ($(\xA,1.95)!0.5!(\xB,1.95)$) {\large$\cdots$};
\node at ($(\xB,1.95)!0.5!(\xC,1.95)$) {\large$\cdots$};

\node[font=\scriptsize, align=center, fill=white, inner sep=1pt] at ($(\xA,3.5)!0.5!(\xB,3.5)$)
  {finite-capacity\\fronthaul};

\coordinate (u1) at (0.9,-0.75);
\coordinate (u2) at (4.3,-0.85);
\coordinate (u3) at (8.4,-0.75);
\coordinate (u4) at (12.0,-0.85);

\foreach \u in {u1,u2,u3,u4}{
  \foreach \l in {1,l}{ \draw[ch] (\u) -- ($(ap\l.south)+(0,-0.55)$); }
}
\draw[ch, offc!60] (u4) -- ($(apL.south)+(0,-0.55)$);
\draw[ch, offc!60] (u3) -- ($(apL.south)+(0,-0.55)$);

\foreach \u/\name in {u1/UE 1, u2/UE $k$, u3/UE $i$, u4/UE $K$}{
  \pic at (\u) {ue};
  \node[below=3mm, font=\scriptsize] at (\u) {\name};
}
\node at ($(u2)!0.5!(u3)$) {\large$\cdots$};
\node[font=\scriptsize, fill=white, inner sep=1pt] at ($(u2)!0.62!($(apl.south)+(0,-0.55)$)$)
  {$\mathbf{h}_{kl}$};

\begin{scope}[shift={(0.1,-2.05)}]
  \draw[ul] (0,0) -- (0.6,0) node[right, text=black, font=\scriptsize]
    {UL: quantized despread pilots ($\tau_p$ tokens/bit)};
  \draw[dl] (6.5,0) -- (7.1,0) node[right, text=black, font=\scriptsize]
    {DL: quantized precoders ($K$ tokens/bit)};
\end{scope}

\end{tikzpicture}
\caption{System model of the centralized cell-free massive MIMO network with
  finite-capacity fronthaul. Each AP quantizes its despread pilot observations
  with $b_l^{\mathrm{UL}}$ bits, while the CPU quantizes the precoding matrix $\mathbf W_l$ of
  AP $l$ with $b_l^{\mathrm{DL}}$ bits and AP $l$ forms its transmit signal locally from the data symbols $\{s_k\}$. The bipartite GNN maps the large-scale
  fading coefficients $\{\beta_{kl}\}$ to a feasible joint bit-allocation under
  the shared budget $\frac{1}{\tau_c}\sum_{l}(\tau_p b_l^{\mathrm{UL}}+K b_l^{\mathrm{DL}})\le B_{\mathrm{tot}}$;
  inactive APs receive zero bits in both directions.}
  \label{fig:system_model}
\end{figure*}
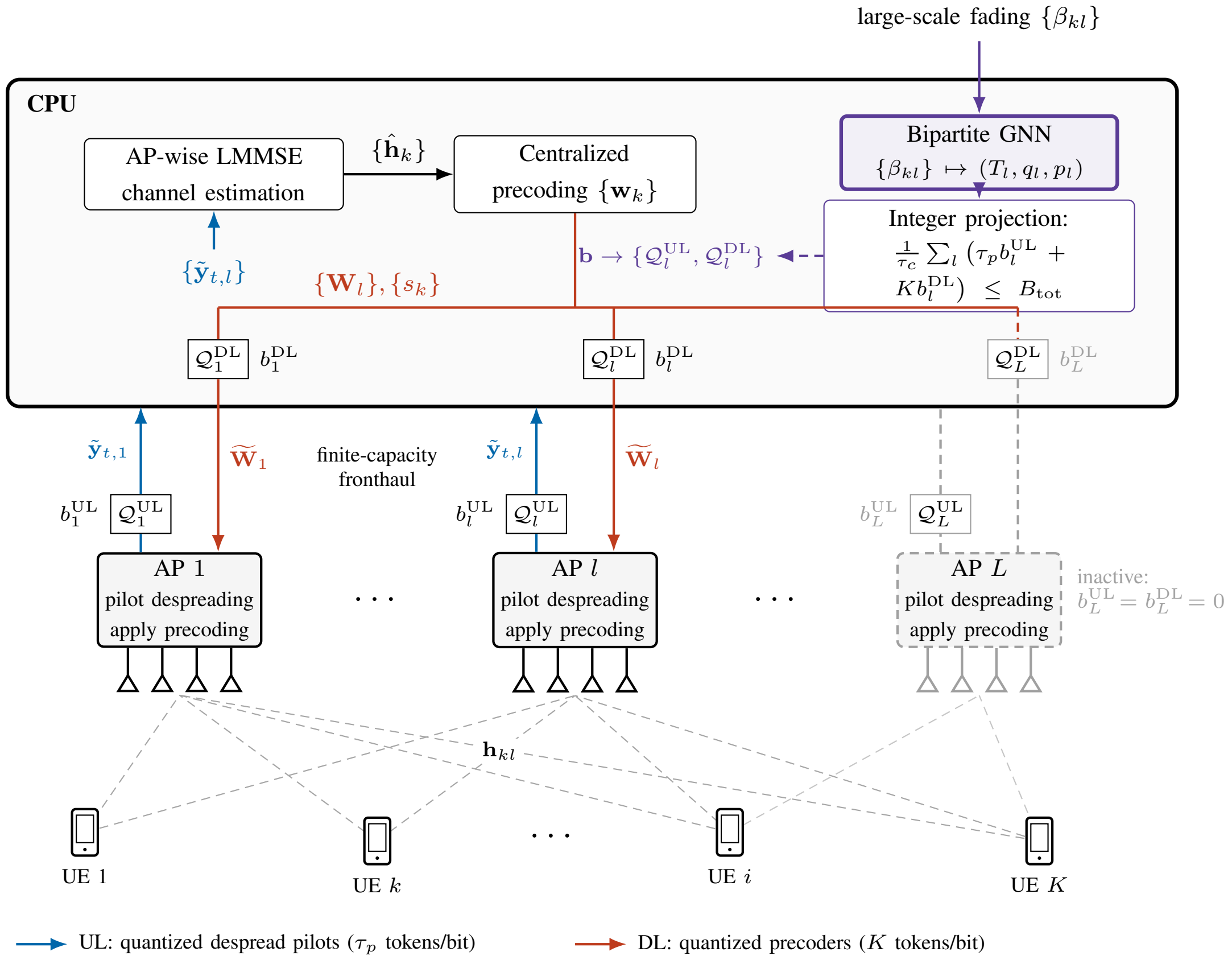

\section{System Model and Problem Formulation}
\label{section:sysmod}
We consider a time-division duplex (TDD) cell-free massive MIMO system with
$L$ distributed APs, each equipped with $N$ antennas, jointly serving $K$
single-antenna UEs, as illustrated in Fig.~\ref{fig:system_model}. The APs are
connected to a CPU through finite-capacity fronthaul links. Each AP quantizes
its despread pilot observations before forwarding them to the CPU, which
performs centralized channel estimation and downlink precoding and then
quantizes the precoding vectors conveyed back to the APs, together with the data symbols.
The channel between UE $k$ and AP $l$ is denoted by
$\mathbf h_{kl}\in\mathbb C^{N}$, and the aggregate channel between UE $k$ and
all APs is
\begin{equation}
    \mathbf h_k =
    \begin{bmatrix}
        \mathbf h_{k1}^{\Ttran} & \cdots & \mathbf h_{kL}^{\Ttran}
    \end{bmatrix}^{\Ttran}
    \in\mathbb C^{LN}.
\end{equation}
We consider spatially correlated Rayleigh fading for AP--UE channels:
\begin{equation}
    \mathbf h_{kl}
    \sim
    \mathcal{CN}(\mathbf 0,\mathbf R_{kl}),
\end{equation}
where $\mathbf R_{kl}\in\mathbb C^{N\times N}$ is the spatial correlation
matrix. The large-scale fading coefficient between UE $k$ and AP $l$ is
\begin{equation}
    \beta_{kl}
    =
    \frac{\mathrm{tr}(\mathbf R_{kl})}{N}.
\end{equation}
We define the aggregate correlation matrix for UE $k$ as 
\begin{equation}
    \mathbf R_k
    =
    \mathrm{diag}
    \left(
        \mathbf R_{k1},\ldots,\mathbf R_{kL}
    \right)
    \in\mathbb C^{LN\times LN}.
\end{equation}
\subsection{Uplink Pilot Transmission and Channel Estimation}
Each coherence block consists of $\tau_c$ channel uses. The first $\tau_p$
channel uses are used for uplink training, while the remaining
$\tau_d=\tau_c-\tau_p$ channel uses are used for downlink data transmission.
A set of $\tau_p$ mutually orthogonal pilot sequences
$\boldsymbol\phi_1,\ldots,\boldsymbol\phi_{\tau_p}\in\mathbb C^{\tau_p}$
is used for training, satisfying
\begin{equation}
    \boldsymbol\phi_t^{\Htran}\boldsymbol\phi_{t'}
    =
    \begin{cases}
        \tau_p, & t=t',\\
        0, & t\neq t'.
    \end{cases}
\end{equation}
UE $k$ is assigned pilot index $t_k\in\{1,\ldots,\tau_p\}$. Since the coherence blocks are finite-sized, more than one UE might be assigned the same pilot. The pilot contamination set of UE $k$ is
defined as
\begin{equation}
    \mathcal P_k
    =
    \{i: t_i=t_k,\ i=1,\ldots,K\}.
\end{equation}
The received pilot matrix at AP $l$ is
\begin{equation}
    \mathbf Y_l^{\rm p}
    =
    \sum_{i=1}^{K}
    \sqrt{\eta_i}
    \mathbf h_{il}
    \boldsymbol\phi_{t_i}^{\Ttran}
    +
    \mathbf N_l^{\rm p},
\end{equation}
where $\eta_i$ is the pilot transmit power of UE $i$, and
$\mathbf N_l^{\rm p}\in\mathbb C^{N\times\tau_p}$ is the noise matrix at AP $l$ with independent and identically distributed (i.i.d.)
$\mathcal{CN}(0,\sigma^2)$ noise samples.
Instead of forwarding the full pilot matrix to the CPU, each AP performs pilot
despreading locally to extract the observation associated with each pilot sequence for subsequent channel estimation. The despread signal corresponding to pilot $t$ at AP $l$
is
\begin{equation}\label{eq:despread}
    \mathbf y_{t,l}
    =
    \mathbf Y_l^{\rm p}\frac{\boldsymbol\phi_t^*}{\sqrt{\tau_p}}
    =
    \sum_{i:t_i=t}
    \sqrt{\eta_i\tau_p}\, \mathbf h_{il}
    +
    \mathbf n_{t,l},
\end{equation}
where
\begin{equation}
    \mathbf n_{t,l}
    =
    \mathbf N_l^{\rm p}\frac{\boldsymbol\phi_t^*}{\sqrt{\tau_p}}
    \sim
    \mathcal{CN}
    \left(
        \mathbf 0,
        \sigma^2\mathbf I_N
    \right).
\end{equation}
This expression shows that the despread pilot observation contains the channels
of all UEs sharing the same pilot.

The despread pilot signal $\mathbf y_{t,l}$ is quantized before being forwarded
to the CPU. Let $b_{l}^{\rm UL}$ denote the number of uplink fronthaul
quantization bits allocated to each despread pilot signal at AP $l$. Since all
$\tau_p$ despread pilot signals of AP $l$ are conveyed over the same fronthaul
link, they are quantized with this common resolution, which also keeps the
fronthaul control overhead low.
Using the additive quantization noise model (AQNM), the quantized despread
pilot signal is modeled as
\begin{equation}
    \widetilde{\mathbf y}_{t,l}
    =
    \alpha_{l}^{\rm UL}
    \mathbf y_{t,l}
    +
    \mathbf q_{t,l}^{\rm UL},
\end{equation}
where
\begin{equation}
    \alpha_{l}^{\rm UL}
    =
    1-\rho_{l}^{\rm UL},
    \qquad
    \rho_{l}^{\rm UL}
    =
    \rho\big(b_{l}^{\rm UL}\big).
\end{equation}
Here, $\rho(b)$ denotes the normalized mean-squared quantization error of the
$b$-bit Lloyd--Max quantizer for a Gaussian input, which is given by
\begin{equation}\label{eq:rho}
    \rho(b)
    =
    \begin{cases}
        1, & b=0,\\
        \bar\rho_b, & b=1,\ldots,5,\\
        c_q2^{-2b}, & b\geq 6,
    \end{cases}
\end{equation}
where $\bar\rho_1=0.3634$, $\bar\rho_2=0.1175$, $\bar\rho_3=0.0345$,
$\bar\rho_4=0.0095$, and $\bar\rho_5=0.0025$ are the tabulated Lloyd--Max
distortion values and $c_q=\pi\sqrt{3}/2$ is the
constant of the high-resolution approximation \cite{Orhan2015a}. Note that $\rho(0)=1$ gives
$\alpha_l^{\rm UL}=0$, i.e., an AP with $b_l^{\rm UL}=0$ forwards no pilot
observation to the CPU.
The covariance matrix of $\mathbf y_{t,l}$ is
\begin{equation}\label{eq:Cy}
    \mathbf C_{y,t,l}
    =
    \mathbb E
    \left\{
        \mathbf y_{t,l}\mathbf y_{t,l}^{\Htran}
    \right\}
    =
    \sum_{i:t_i=t}
    \eta_i\tau_p \mathbf R_{il}
    +
    \sigma^2\mathbf I_N.
\end{equation}
The AQNM quantization noise covariance is \cite{Orhan2015a}
\begin{equation}\label{eq:Cq_ul}
    \mathbf C_{q,t,l}^{\rm UL}
    =
    \alpha_{l}^{\rm UL}
    \left(
        1-\alpha_{l}^{\rm UL}
    \right)
    \mathrm{diag}
    \left(
        \mathbf C_{y,t,l}
    \right).
\end{equation}
\begin{remark}\label{remark:aqnm}
The AQNM is an approximate model of the actual scalar quantizer. The
uncorrelatedness of $\mathbf q_{t,l}^{\rm UL}$ and $\mathbf y_{t,l}$ is not
the source of the approximation, since the Bussgang decomposition of a scalar quantizer with Gaussian input also yields a distortion term that is
uncorrelated with the quantizer input. The approximation lies in that the
AQNM treats $\mathbf q_{t,l}^{\rm UL}$ as a noise term that is independent of
$\mathbf y_{t,l}$ and has uncorrelated entries, with the diagonal covariance
matrix in \eqref{eq:Cq_ul}, and this is also how it enters the surrogate rate
evaluation in Section~\ref{section:greedy}. However, the actual quantization
distortion is a deterministic function of the quantizer input and, hence, it
is statistically dependent on it. Furthermore, its entries are correlated
across the antennas when the entries of $\mathbf y_{t,l}$ are spatially
correlated. On the other hand, the AQNM provides a tractable surrogate model whose parameters can
be directly updated for every candidate bit-allocation. In this paper, we
utilize the AQNM, together with its downlink counterpart in the next
subsection, only for the bit-allocation optimization in
Section~\ref{section:greedy} and the GNN training in
Section~\ref{section:proposed}. The spectral efficiencies reported in
Section~\ref{section:results} are not based on the AQNM. Instead, they are
obtained with the actual Lloyd--Max quantizers 
as described in Section~\ref{section:Bussgang}.
\end{remark}

Although downlink precoding is performed centrally at the CPU, the channel
estimation problem decomposes across APs since the channels and receiver noises
are independent across APs. Therefore, after receiving the quantized despread
pilot signals from all APs, the CPU can compute the linear minimum mean-squared error (LMMSE) channel estimates
separately for each AP.
For UE $k$ and AP $l$, the CPU estimates $\mathbf h_{kl}$ from
$\widetilde{\mathbf y}_{t_k,l}$. Since $\alpha_{l}^{\rm UL}$ is real-valued,
the covariance matrix of the quantized despread observation $\widetilde{\mathbf y}_{t_k,l}$ is
\begin{equation}
    \mathbf \Psi_{t_k,l}^{\rm UL}
    =
    \left(
        \alpha_{l}^{\rm UL}
    \right)^2
    \mathbf C_{y,t_k,l}
    +
    \mathbf C_{q,t_k,l}^{\rm UL},
\end{equation}
where the sum in \eqref{eq:Cy} is taken over
$i\in\mathcal P_k$.
The cross-covariance between $\mathbf h_{kl}$ and
$\widetilde{\mathbf y}_{t_k,l}$ is
\begin{equation}
    \mathbb E
    \left\{
        \mathbf h_{kl}
        \widetilde{\mathbf y}_{t_k,l}^{\Htran}
    \right\}
    =
    \sqrt{\eta_k\tau_p}\,
    \alpha_{l}^{\rm UL}
    \mathbf R_{kl}.
\end{equation}
Hence, the AP-wise LMMSE estimate is
\begin{equation}
    \widehat{\mathbf h}_{kl}
    =
    \sqrt{\eta_k\tau_p}\,
    \alpha_{l}^{\rm UL}
    \mathbf R_{kl}
    \left(
        \mathbf \Psi_{t_k,l}^{\rm UL}
    \right)^{-1}
    \widetilde{\mathbf y}_{t_k,l}.
\end{equation}
The estimation error
$\widetilde{\mathbf h}_{kl}=\mathbf h_{kl}-\widehat{\mathbf h}_{kl}$
has the covariance matrix
\begin{equation}
    \mathbf C_{kl}
    =
    \mathbf R_{kl}
    -
    \eta_k\tau_p
    \left(
        \alpha_{l}^{\rm UL}
    \right)^2
    \mathbf R_{kl}
    \left(
        \mathbf \Psi_{t_k,l}^{\rm UL}
    \right)^{-1}
    \mathbf R_{kl},
\end{equation}
and the covariance matrix of the channel estimate is
$\widehat{\mathbf R}_{kl}=\mathbf R_{kl}-\mathbf C_{kl}$.
After computing the AP-wise channel estimates, the CPU forms the aggregate
channel estimate used for centralized precoding as
\begin{equation}
    \widehat{\mathbf h}_{k}
    =
    \begin{bmatrix}
        \widehat{\mathbf h}_{k1}^{\Ttran} &
        \cdots &
        \widehat{\mathbf h}_{kL}^{\Ttran}
    \end{bmatrix}^{\Ttran}
    \in\mathbb C^{LN}.
\end{equation}
Similarly, the aggregate estimation error covariance is
$\mathbf C_k=\mathrm{diag}\left(\mathbf C_{k1},\ldots,\mathbf C_{kL}\right)$
and the aggregate channel estimate covariance is
$\widehat{\mathbf R}_k=\mathrm{diag}\left(\widehat{\mathbf R}_{k1},\ldots,\widehat{\mathbf R}_{kL}\right)$.
\subsection{Centralized Downlink Precoding}
Based on $\{\widehat{\mathbf h}_k\}_{k=1}^K$, the CPU designs centralized
precoding vectors
\begin{equation}
    \mathbf w_k
    =
    \begin{bmatrix}
        \mathbf w_{k1}^{\Ttran} &
        \cdots &
        \mathbf w_{kL}^{\Ttran}
    \end{bmatrix}^{\Ttran}
    \in\mathbb C^{LN},
    \qquad k=1,\ldots,K,
\end{equation}
where $\mathbf w_{kl}\in\mathbb C^N$ is the precoding subvector corresponding
to UE $k$ at AP $l$.
Without fronthaul quantization, the aggregate downlink signal would be
\begin{equation}
    \mathbf x
    =
    \sum_{k=1}^{K}
    \mathbf w_k s_k
    =
    \begin{bmatrix}
        \mathbf x_1^{\Ttran} &
        \cdots &
        \mathbf x_L^{\Ttran}
    \end{bmatrix}^{\Ttran}
    \in\mathbb C^{LN},
\end{equation}
where $s_k$ is the information symbol intended for UE $k$ with
$\mathbb E\{|s_k|^2\}=1$.
The signal to be sent by AP $l$ is therefore
\begin{equation}
    \mathbf x_l
    =
    \sum_{k=1}^{K}
    \mathbf w_{kl}s_k
    \in\mathbb C^N,
\end{equation}
with the covariance matrix
\begin{equation}
    \mathbf C_{x,l}
    =
    \mathbb E
    \left\{
        \mathbf x_l\mathbf x_l^{\Htran}
    \right\}
    =
    \sum_{k=1}^{K}
    \mathbf w_{kl}\mathbf w_{kl}^{\Htran}.
\end{equation}

Instead of forwarding the precoded samples $\mathbf x_l$, which would have to be conveyed for each of the $\tau_d$ downlink channel uses of a coherence block, the CPU sends to AP $l$ the precoding matrix
\begin{equation}
    \mathbf W_l
    =
    \begin{bmatrix}
        \mathbf w_{1l} & \cdots & \mathbf w_{Kl}
    \end{bmatrix}
    \in\mathbb C^{N\times K}
\end{equation}
only once per coherence block, together with the data symbols $\{s_k\}$, and AP $l$ forms its transmit signal locally. The data symbols are conveyed digitally as coded information bits; hence, their fronthaul load does not depend on the quantization resolutions considered in this paper.\footnote{The data payload is a fixed overhead that is independent of the bit-allocation and is therefore not included in the fronthaul budget in \eqref{eq:budget}.} Let $b_l^{\rm DL}$ denote the number of downlink fronthaul quantization bits allocated to the entries of $\mathbf W_l$. Using AQNM, the quantized precoding vector of UE $k$ at AP $l$ is
\begin{equation}
    {\widetilde{\mathbf w}_{kl}
    =
    \alpha_l^{\rm DL}\mathbf w_{kl}
    +
    \mathbf q_{kl}^{\rm DL},}
\end{equation}
where
\begin{equation}
    \alpha_l^{\rm DL}
    =
    1-\rho_l^{\rm DL},
    \qquad
    \rho_l^{\rm DL}
    =
    \rho\big(b_l^{\rm DL}\big),
\end{equation}
with $\rho(\cdot)$ given in \eqref{eq:rho}.
The quantization noise covariance is
\begin{equation}
    {\mathbf C_{q,kl}^{\rm DL}
    =
    \alpha_l^{\rm DL}
    \left(
        1-\alpha_l^{\rm DL}
    \right)
    \frac{1}{K}
    \mathrm{diag}
    \left(
        \mathbf C_{x,l}
    \right).}
\end{equation}
Here, all precoder entries at a given antenna of AP $l$ are quantized with a common scale determined by their average power over the $K$ UEs, which requires only $N$ real scalars of side information per AP and coherence block.
Since the quantized precoders are kept fixed over the $\tau_d$ downlink channel uses, $\mathbf q_{kl}^{\rm DL}$ acts as a perturbation of the precoding vector rather than as a noise term that changes from one channel use to another. AP $l$ transmits
\begin{equation}
    \widetilde{\mathbf x}_l
    =
    \sum_{k=1}^{K}\widetilde{\mathbf w}_{kl}s_k
    =
    \alpha_l^{\rm DL}\mathbf x_l
    +
    \sum_{k=1}^{K}\mathbf q_{kl}^{\rm DL}s_k .
\end{equation}
Stacking the quantized precoding vectors of UE $k$ from all APs gives
\begin{equation}
    {\widetilde{\mathbf w}_k
    =
    \mathbf A^{\rm DL}\mathbf w_k
    +
    \mathbf q_k^{\rm DL},}
\end{equation}
where
\begin{equation}
    \mathbf A^{\rm DL}
    =
    \mathrm{diag}
    \left(
        \alpha_1^{\rm DL}\mathbf I_N,\ldots,
        \alpha_L^{\rm DL}\mathbf I_N
    \right),
\end{equation}
and $\mathbf q_k^{\rm DL}$ has the block-diagonal covariance $\mathbf C_{q,k}^{\rm DL}=\mathrm{diag}(\mathbf C_{q,k1}^{\rm DL},\ldots,\mathbf C_{q,kL}^{\rm DL})$. The transmitted signal is $\widetilde{\mathbf x}=\mathbf A^{\rm DL}\mathbf x+\sum_{k}\mathbf q_k^{\rm DL}s_k$, where the aggregate quantization noise $\sum_{k}\mathbf q_k^{\rm DL}s_k$ has the covariance
\begin{equation}
    \mathbf C_q^{\rm DL}
    =
    {\sum_{k=1}^{K}\mathbf C_{q,k}^{\rm DL}
    =}
    \mathrm{diag}
    \left(
        \mathbf C_{q,1}^{\rm DL},\ldots,
        \mathbf C_{q,L}^{\rm DL}
    \right),
\end{equation}
with $\mathbf C_{q,l}^{\rm DL}=\alpha_l^{\rm DL}(1-\alpha_l^{\rm DL})\mathrm{diag}(\mathbf C_{x,l})$.

Since AP $l$ knows $\widetilde{\mathbf W}_l$, the transmit power constraint is imposed on the quantized precoders as
\begin{equation}
    {\mathbb E
    \left\{
        \|\widetilde{\mathbf x}_l\|^2
    \right\}
    =
    \sum_{k=1}^{K}\|\widetilde{\mathbf w}_{kl}\|^2}
    \leq
    P_{\max},
    \qquad l=1,\ldots,L, 
\end{equation}
where the expectation is with respect to the data symbols and $P_{\max}$ is the maximum power available at each AP.
\subsection{Problem Formulation}
We collect all bit-allocation variables in the vector  $\mathbf b$ as 
\begin{equation}
    \mathbf b
    =
    \left[
        \{b_{l}^{\rm UL}\}_{l},
        \{b_l^{\rm DL}\}_{l}
    \right]^{\Ttran}
    \in
    \mathcal B^{M_b},
    \qquad
    M_b=2L,
\end{equation}
where $\mathcal B=\{0,1,\ldots,b_{\max}\}$ is the set of feasible integer bit
values, with $b_{\max}$ denoting the maximum quantization resolution. We consider a single total fronthaul bit budget shared by uplink pilot
quantization and downlink precoder quantization. In each coherence block, AP $l$ forwards $\tau_p$ despread pilot vectors of dimension $N$, each quantized with $b_l^{\rm UL}$ bits per real dimension, which requires $2N\tau_p b_l^{\rm UL}$ bits. The CPU sends $K$ precoding vectors of dimension $N$ to AP $l$, which requires $2NKb_l^{\rm DL}$ bits. Normalizing the total load by $2N\tau_c$, we obtain the fronthaul budget constraint
\begin{equation}\label{eq:budget}
    {\frac{1}{\tau_c}
    \sum_{l=1}^{L}
    \left(
    \tau_p\, b_{l}^{\rm UL}
    +
    K\, b_l^{\rm DL}
    \right)
    \leq
    B_{\rm tot},}
\end{equation}
where $B_{\rm tot}$ is the total fronthaul budget in bits per channel use and per real antenna dimension. Hence, one uplink bit consumes $\tau_p$ budget tokens and one downlink bit consumes $K$ tokens out of the $\tau_cB_{\rm tot}$ available tokens. For comparison, forwarding the precoded samples $\{\mathbf x_l\}$ instead of the precoders would make one downlink bit consume $\tau_d$ tokens, which is much larger than $K$ in typical settings, e.g., $\tau_d=192$ versus $K=16$ in Section~\ref{section:results}. In that case, the downlink load would dominate the shared budget and the uplink--downlink trade-off would largely disappear.
To guarantee a minimum quantization resolution at every AP that is served
through the fronthaul, while still allowing an AP to be switched off when
quantizing its link is not worthwhile, we impose for each $l=1,\ldots,L$
\begin{equation}\label{eq:bmin}
    b_{l}^{\rm UL}=b_l^{\rm DL}=0
    \quad \mathrm{or} \quad
    b_{l}^{\rm UL}\geq b_{\min},\ b_l^{\rm DL}\geq b_{\min},
\end{equation}
where $b_{\min}\geq 1$. We refer to allocations satisfying \eqref{eq:bmin}
as \emph{AP-consistent}.

The uplink bits affect the downlink performance implicitly through the quality
of the channel estimates used for centralized precoding, while the downlink
bits affect it explicitly through the quantization distortion of the precoding vectors used at the APs. The joint bit-allocation problem is
\begin{subequations}\label{eq:problem}
\begin{align}
    \underset{\mathbf b\in\mathcal B^{M_b}}{\mathrm{maximize}}
    \quad
    &
    \mathbb E
    \left\{
        \sum_{k=1}^{K}
        R_k(\mathbf b)
    \right\}
    \\
    \mathrm{subject~to}
    \quad
    &
    \eqref{eq:budget},\
     \eqref{eq:bmin}, 
\end{align}
\end{subequations}
where $R_k(\mathbf b)$ is the downlink rate of UE $k$ under the bit-allocation
$\mathbf b$ and the expectation is with respect to the small-scale fading.
Problem \eqref{eq:problem} is a combinatorial integer program without a
closed-form objective, and an exhaustive search over $\mathcal B^{M_b}$ is
intractable even for moderate $L$. Moreover, the bit-allocation
must be recomputed whenever the large-scale propagation conditions change. To
address both issues, in the next section we first develop a greedy algorithm
that finds a locally optimal solution to \eqref{eq:problem} for a given
network setup, and then train a GNN in a supervised manner to learn the mapping from the large-scale fading coefficients to the greedy bit-allocations, enabling fast real-time inference.

As discussed in Remark~\ref{remark:aqnm}, the AQNM is adopted only during the
bit-allocation optimization and the GNN training.
\section{Greedy Bit-Allocation for Label Generation }
\label{section:greedy}
In this section, we develop the greedy algorithm that is used to generate the labels for the supervised GNN training in Section~\ref{section:proposed}. Since the objective of \eqref{eq:problem} has no closed-form expression, we first describe in Section~\ref{sec:aqnm_sample_avg} how it is evaluated for a given bit-allocation using the AQNM and a fixed set of channel realizations. Building on this evaluation, Section~\ref{sec:greedy_label_generation} presents the greedy search over the feasible allocations, its initialization, and the resulting training dataset.
\subsection{AQNM-Based Sample-Average Rate Evaluation}
\label{sec:aqnm_sample_avg}
For each large-scale fading setup $\{\mathbf R_{kl}\}$, we generate a fixed
set of $X$ independent small-scale fading realizations according to
\begin{equation}
    \mathbf h_{kl}^{(x)}
    =
    \mathbf R_{kl}^{1/2}
    \mathbf g_{kl}^{(x)},
    \qquad
    \mathbf g_{kl}^{(x)}
    \sim
    \mathcal{CN}(\mathbf 0,\mathbf I_N),
    \qquad x=1,\ldots,X.
\end{equation}
The same set of channel realizations is used for all candidate bit-allocations
in order to obtain a fair sample-average comparison.
For a given bit-allocation vector $\mathbf b$, we first recover the uplink
bits $\{b_{l}^{\rm UL}\}$ and downlink bits $\{b_l^{\rm DL}\}$. The uplink
bits determine the AQNM coefficients used for pilot quantization and,
consequently, the LMMSE channel estimation quality. The LMMSE filters
$\sqrt{\eta_k\tau_p}\,\alpha_{l}^{\rm UL}\mathbf R_{kl}
(\mathbf \Psi_{t_k,l}^{\rm UL})^{-1}$ depend only on the statistics and can be
computed once for each candidate bit-allocation. Then, for each realization
$x=1,\ldots,X$, the quantized despread pilot observations are generated
according to the AQNM, the LMMSE channel estimates are computed, and the
centralized downlink precoding vectors $\mathbf w_k^{(x)}$ are designed based
on these estimates. The downlink bits determine the AQNM coefficients and
quantization-noise covariance of the quantized precoding vectors. The downlink
rate is then evaluated using the actual instantaneous channel realization. For
realization $x$, the instantaneous signal-to-interference-plus-noise ratio (SINR) of UE $k$ is computed as in
\eqref{eq:aqnm_sinr}, shown at the top of the next page.
\begin{figure*}[!t]
\centering
\begin{equation}
\gamma_k^{(x)}(\mathbf b)
=
\frac{
\left|
\left(\mathbf h_k^{(x)}\right)^{\Htran}
\mathbf A^{\rm DL}(\mathbf b)
\mathbf w_k^{(x)}
\right|^2
}{
\sum_{i\neq k}
\left|
\left(\mathbf h_k^{(x)}\right)^{\Htran}
\mathbf A^{\rm DL}(\mathbf b)
\mathbf w_i^{(x)}
\right|^2
+
\left(\mathbf h_k^{(x)}\right)^{\Htran}
\mathbf C_q^{{\rm DL},(x)}(\mathbf b)
\mathbf h_k^{(x)}
+
\sigma^2
}
\label{eq:aqnm_sinr}
\end{equation}
\vspace{-0.5em}
\noindent\rule{\textwidth}{0.4pt}
\vspace{-1.0em}
\end{figure*}
Here, $\mathbf h_k^{(x)}$ is the actual aggregate channel of UE $k$ in
realization $x$, and $\mathbf w_k^{(x)}$ is the precoding vector computed from
the corresponding quantized pilot-based channel estimates, and $\mathbf C_q^{{\rm DL},(x)}(\mathbf b)=\sum_{i=1}^{K}\mathbf C_{q,i}^{{\rm DL},(x)}$ collects the AQNM noise of all quantized precoders in realization $x$. The sample-average
sum rate for the candidate bit-allocation is then
\begin{equation}\label{eq:avg_rate}
\bar R_{\rm sum}(\mathbf b)
=
\frac{1}{X}
\sum_{x=1}^{X}
\sum_{k=1}^{K}
\frac{\tau_d}{\tau_c}
\log_2
\left(
1+\gamma_k^{(x)}(\mathbf b)
\right).
\end{equation}
\subsection{Greedy Label Generation by Pairwise Bit Exchange}
\label{sec:greedy_label_generation}
We use $b_{\min}=1$ in \eqref{eq:bmin}. Since a silent AP is excluded from the centralized precoder and an
AP without channel estimates receives no precoding power, we restrict the search
to the AP-consistent allocations in \eqref{eq:bmin}, in which every AP is either fully active
($b_l^{\rm UL}\ge 1$ and $b_l^{\rm DL}\ge 1$) or fully off
($b_l^{\rm UL}=b_l^{\rm DL}=0$).

\begin{algorithm}[t]
\caption{Greedy Label Generation}
\label{alg:label_generation_compact}
\small
\algrenewcommand\algorithmicindent{0.8em}
\begin{algorithmic}[1]
\Require Spatial correlation matrices $\{\mathbf R_{kl}\}$,
$B_{\rm tot}$, $b_{\min}$, number of realizations $X$
\Ensure Locally optimized bit-allocation $\mathbf b^\star$

\LongState{Set $M_b=2L$,
$\mathbf b=[\{b_l^{\rm UL}\}_{l},
\{b_l^{\rm DL}\}_{l}]^{\Ttran}$, and
$\mathbf w=[\tau_p\mathbf 1_L^{\Ttran},
K\mathbf 1_L^{\Ttran}]^{\Ttran}$.}

\LongState{Generate fixed channels
$\mathbf h_{kl}^{(x)}=\mathbf R_{kl}^{1/2}
\mathbf g_{kl}^{(x)}$,
$\mathbf g_{kl}^{(x)}\sim\mathcal{CN}
(\mathbf 0,\mathbf I_N)$,
$x=1,\ldots,X$.}

\LongState{Let $\mathcal F$ denote the set of
AP-consistent allocations in $\mathcal B^{M_b}$
satisfying \eqref{eq:bmin} and
$\sum_{i=1}^{M_b}w_i b_i\le\tau_cB_{\rm tot}$.}

\LongState{Initialize $\mathbf b\in\mathcal F$ and set
$R_{\rm cur}\gets\bar R_{\rm sum}(\mathbf b)$
in \eqref{eq:avg_rate}.}

\Repeat
    \State $\texttt{changed}\gets\texttt{false}$.
    \For{$i=1,\ldots,M_b-1$}
        \For{$j=i+1,\ldots,M_b$}
            \LongState{$\mathcal C\gets\emptyset$,
            $g_{ij}\gets\gcd(w_i,w_j)$,
            $d_i\gets w_j/g_{ij}$,
            $d_j\gets w_i/g_{ij}$.}

            \LongState{\textbf{if}
            $\mathbf b-d_i\mathbf e_i+d_j\mathbf e_j
            \in\mathcal F$ \textbf{then}
            $\mathcal C\gets\mathcal C\cup
            \{\mathbf b-d_i\mathbf e_i+d_j\mathbf e_j\}$.}

            \LongState{\textbf{if}
            $\mathbf b-d_j\mathbf e_j+d_i\mathbf e_i
            \in\mathcal F$ \textbf{then}
            $\mathcal C\gets\mathcal C\cup
            \{\mathbf b-d_j\mathbf e_j+d_i\mathbf e_i\}$.}

            \If{$\mathcal C\neq\emptyset$}
                \LongState{$\mathbf b_{\rm best}\gets
                \arg\max_{\tilde{\mathbf b}\in\mathcal C}
                \bar R_{\rm sum}(\tilde{\mathbf b})$,
                $R_{\rm best}\gets
                \bar R_{\rm sum}(\mathbf b_{\rm best})$.}

                \LongState{\textbf{if}
                $R_{\rm best}>R_{\rm cur}$ \textbf{then}
                $\mathbf b\gets\mathbf b_{\rm best}$,
                $R_{\rm cur}\gets R_{\rm best}$,
                $\texttt{changed}\gets\texttt{true}$.}
            \EndIf
        \EndFor
    \EndFor

    \For{$i=1,\ldots,M_b$}
        \LongState{\textbf{if}
        $\mathbf b+\mathbf e_i\in\mathcal F$ and
        $\bar R_{\rm sum}(\mathbf b+\mathbf e_i)
        >R_{\rm cur}$ \textbf{then}
        $\mathbf b\gets\mathbf b+\mathbf e_i$,
        $R_{\rm cur}\gets\bar R_{\rm sum}(\mathbf b)$,
        $\texttt{changed}\gets\texttt{true}$.}
    \EndFor

    \For{$l=1,\ldots,L$}
        \LongState{$\widetilde{\mathbf b}\gets
        \Call{APswitch}{\mathbf b,l}$,
        $\widetilde R\gets
        \bar R_{\rm sum}(\widetilde{\mathbf b})$.}

        \LongState{\textbf{if}
        $\widetilde R>R_{\rm cur}$ \textbf{then}
        $\mathbf b\gets\widetilde{\mathbf b}$,
        $R_{\rm cur}\gets\widetilde R$,
        $\texttt{changed}\gets\texttt{true}$.}
    \EndFor
\Until{$\texttt{changed}=\texttt{false}$}

\State $\mathbf b^\star\gets\mathbf b$;
\Return $\mathbf b^\star$.
\end{algorithmic}
\end{algorithm}

The proposed greedy procedure is summarized in
Algorithm~\ref{alg:label_generation_compact}. Starting from a feasible integer
bit-allocation, each iteration
first performs a sweep over all pairs of bit-allocation variables. For each
pair, the algorithm constructs candidates that transfer fronthaul budget in
both directions while preserving the weighted budget constraint. In
particular, since one uplink bit consumes $\tau_p$ tokens whereas one downlink
bit consumes $K$ tokens, an uplink--downlink exchange trades $K/g$
uplink bits for $\tau_p/g$ downlink bits, or vice versa, where $g=\gcd(\tau_p,K)$; for the parameters in Section~\ref{section:results}, two uplink bits are traded for one downlink bit. Uplink--uplink and
downlink--downlink exchanges reduce to ordinary one-bit transfers when the
exchange step is one. Only candidates satisfying the bit limits and the
AP-consistency constraint are considered, and a candidate is accepted
immediately if it strictly increases the sample-average sum rate
$\bar R_{\mathrm{sum}}(\mathbf b)$. Since the budget in \eqref{eq:budget} is an inequality, the initial rounding and the AP activation/deactivation moves, which are performed at the end of each iteration as described below, may leave some tokens unused, e.g., when the released tokens cannot be spent because the remaining active resolutions have reached their maximum. The pairwise exchanges cannot recover these tokens because they preserve the number of used tokens. Therefore, in each iteration, after the pairwise sweep, the algorithm also tests the single-bit increments $\mathbf b+\mathbf e_i$ that fit into the remaining budget. In this way, the tokens left unused by the initial rounding or by the AP moves of the previous iteration can be spent.

After completing the pairwise-exchange and single-increment sweeps, the algorithm examines an
additional activation or deactivation move for each AP. This is done by $\textsc{APswitch}(\mathbf b,l)$ in step $22$ of Algorithm \ref{alg:label_generation_compact} which generates a candidate allocation by changing
the activity status of AP $l$. If AP $l$ is active,
the algorithm forms a candidate in which both $b_l^{\mathrm{UL}}$ and
$b_l^{\mathrm{DL}}$ are set to zero, thereby releasing
$\tau_p b_l^{\mathrm{UL}}+K b_l^{\mathrm{DL}}$ tokens. These tokens are then
redistributed among the remaining active APs without exceeding the weighted budget and while preserving the AP-consistency constraint, and the last active AP is never switched off. If AP
$l$ is inactive, the algorithm attempts to activate it with the minimum
AP-consistent allocation
$(b_l^{\mathrm{UL}},b_l^{\mathrm{DL}})=(1,1)$. The required
$\tau_p+K$ tokens are taken from eligible variables of the currently active
APs, and any excess tokens resulting from the unequal uplink and downlink token
weights are redistributed. An infeasible activation or deactivation candidate
is discarded, whereas a feasible candidate is accepted immediately if it
improves $\bar R_{\mathrm{sum}}(\mathbf b)$.

For every candidate, the sample-average rate is re-evaluated so that changes in
the uplink bits are reflected through channel estimation, while changes in the
downlink bits are reflected through fronthaul quantization during data
transmission. Since every accepted move strictly increases
$\bar R_{\mathrm{sum}}(\mathbf b)$ and the feasible set is finite, the
algorithm converges in a finite number of iterations to a locally optimal
allocation with respect to the combined neighborhood of weighted pairwise
exchanges, single-bit increments, and AP activation/deactivation moves.

For initialization, AP $l$ is assigned uplink and downlink importance scores
based on its aggregate large-scale channel strength as
\begin{equation}
\begin{aligned}
\label{eq:aggregate_strengths}
s_l^{\rm UL}
&=
\log_2\left(1+\frac{\tau_p}{\sigma^2}\sum_{k=1}^{K}\eta_k \beta_{kl}\right),
\\
s_l^{\rm DL}
&= 
\log_2\left(1+\frac{P_{\max}}{\sigma^2}\sum_{k=1}^{K}\beta_{kl}\right).
\end{aligned}
\end{equation}
The corresponding fractions of the total fronthaul token budget are then calculated as 
\begin{equation}
\label{eq:token_budget_fraction}
f_l^{\rm UL}
=
\frac{s_l^{\rm UL}}
{\sum_{j=1}^{L}\left(s_j^{\rm UL}+s_j^{\rm DL}\right)},
~
f_l^{\rm DL}
=
\frac{s_l^{\rm DL}}
{\sum_{j=1}^{L}\left(s_j^{\rm UL}+s_j^{\rm DL}\right)}.
\end{equation}
Hence, the initial continuous bit-allocations are
\begin{equation}
\label{eq:initial_bit_allocation}
{\bar b_l^{\rm UL}
=
\frac{\tau_cB_{\rm tot}f_l^{\rm UL}}{\tau_p},
\qquad
\bar b_l^{\rm DL}
=
\frac{\tau_cB_{\rm tot}f_l^{\rm DL}}{K}.}
\end{equation}
These values are then mapped to feasible integer resolutions by flooring and distributing the remaining tokens according to the largest fractional remainders\footnote{For a continuous allocation $\bar b_i$, its fractional remainder is $r_i=\bar b_i-\lfloor\bar b_i\rfloor$. After flooring all allocations, the remaining tokens are assigned, whenever feasible, to the entries with the largest values of $r_i$, while accounting for the token cost of each added bit. 
For example, an entry with the continuous allocation $\bar b_i=3.8$ (floored to $3$) receives an additional bit before an entry with $\bar b_j=2.3$ (floored to $2$), because $r_i=0.8>r_j=0.3$.}, while satisfying the bit limits, the total weighted budget, and the AP-consistency constraint. 

Running Algorithm~\ref{alg:label_generation_compact} for a set of $T_{\rm tr}$
independent large-scale fading setups produces a labeled training dataset
\begin{equation}
    \mathcal D
    =
    \left\{
        \left(
            \boldsymbol\beta^{(n)},
            \mathbf b^{\star(n)}
        \right)
    \right\}_{n=1}^{T_{\rm tr}},
\end{equation}
where $\boldsymbol\beta^{(n)}$ collects the large-scale fading coefficients of
setup $n$.

\section{Proposed GNN-Based Bit-Allocation}
\label{section:proposed}
We learn the mapping from the large-scale fading coefficients to the greedy bit
allocation with a permutation-equivariant GNN that
operates on the bipartite UE--AP graph. Each UE $k$ and each AP $l$ is a node,
and every UE--AP pair $(k,l)$ is an edge carrying the scalar input feature
\begin{equation}
    e_{kl}^{(0)}=\frac{\log_{10}\beta_{kl}-\mu_\beta}{\sigma_\beta},
\end{equation}
where $\mu_\beta$ and $\sigma_\beta$ are the mean and standard deviation of
$\log_{10}\beta_{kl}$ over the training set. Using the large-scale fading on a
logarithmic scale as the sole input makes the allocation depend on the geometry
through the quantity that governs both channel estimation and precoding, while
being equivariant to any relabeling of the UEs and the APs.

\subsubsection*{Message Passing} The scalar edge features are embedded using a multilayer perceptron (MLP) as
$\mathbf e_{kl}\leftarrow\mathrm{MLP}_{\rm e}(e_{kl}^{(0)})\in\mathbb R^{H}$, and
the AP and UE embeddings are initialized as
$\mathbf a_l=\tfrac1K\sum_{k}\mathbf e_{kl}$ and
$\mathbf u_k=\tfrac1L\sum_{l}\mathbf e_{kl}$. For $r=1,\ldots,R$ rounds,
\begin{align}
    \mathbf e_{kl}&\leftarrow\mathbf e_{kl}
        +\mathrm{MLP}_{\rm e}^{(r)}\!\big([\mathbf e_{kl},\mathbf u_k,\mathbf a_l]\big),\\
    \mathbf a_l&\leftarrow\mathbf a_l
        +\mathrm{MLP}_{\rm a}^{(r)}\!\big([\textstyle\frac1K\sum_k\mathbf e_{kl},\,\max_k\mathbf e_{kl},\,\mathbf a_l]\big),\\
    \mathbf u_k&\leftarrow\mathbf u_k
        +\mathrm{MLP}_{\rm u}^{(r)}\!\big([\textstyle\frac1L\sum_l\mathbf e_{kl},\,\max_l\mathbf e_{kl},\,\mathbf u_k]\big),
\end{align}
where $[\,\cdot\,]$ denotes concatenation and the mean/max aggregations run over
the neighbors of a node on the bipartite graph. All MLPs are shared across
edges, APs, and UEs, so the network is permutation equivariant and generalizes
to different numbers of UEs and APs without retraining.

\subsubsection*{Per-AP Outputs} After $R$ rounds, each AP embedding $\mathbf a_l$ is read
out by a shared MLP into three scalars, activated as
\begin{equation}
    T_l=\mathrm{softmax}_l\!\big(s_l^{\rm tot}\big),\quad
    q_l=\sigma\!\big(s_l^{\rm rat}\big),\quad
    p_l=\sigma\!\big(s_l^{\rm act}\big),
\end{equation}
with $(s_l^{\rm tot},s_l^{\rm rat},s_l^{\rm act})=\mathrm{MLP}_{\rm out}(\mathbf a_l)$
and $\sigma(\cdot)$ the logistic function. The three heads have a direct
operational meaning under the weighted budget \eqref{eq:budget}:
(i)~$T_l$ is AP $l$'s share of the token budget $\tau_cB_{\rm tot}$; (ii)~$q_l$ is the
fraction of AP $l$'s tokens spent on downlink fronthaul quantization, so
$(1-q_l)$ is spent on uplink pilot quantization; and (iii)~$p_l$ is the
probability that AP $l$ is active. A hard gate switches AP $l$ fully off when
$p_l<\tfrac12$, forcing both its uplink and downlink bits to zero. Supervising
the on/off decision with a dedicated head, instead of thresholding $T_l$,
removes the sensitivity of the gate to small errors in $T_l$ at tight budgets.

\subsubsection*{Projection} The continuous outputs are mapped to a feasible integer bit vector by first
setting the token shares of APs with $p_l<1/2$ to zero and renormalizing the
remaining shares over the active APs so that they sum to one. The resulting
share of each active AP is then divided between uplink and downlink by
allocating $T_l(1-q_l)\tau_cB_{\rm tot}$ and $T_lq_l\tau_cB_{\rm tot}$ tokens,
respectively. These token allocations are converted to integer bit resolutions
by flooring and distributing the residual tokens according to the largest
fractional remainders under \eqref{eq:budget}, while enforcing the
AP-consistency constraint.

\subsubsection*{Targets and Loss} From each greedy label $\mathbf b^\star$, we form the
per-AP targets
\begin{equation}
    {T_l^\star=\frac{c_l^\star}{\sum_{j=1}^{L}c_j^\star},~
    q_l^\star=\frac{K b_l^{\mathrm{DL}\star}}{c_l^\star}},~
    a_l^\star=\mathbf 1\{b_l^{\mathrm{UL}\star}>0\},
\end{equation}where $c_l^\star=\tau_p b_l^{\mathrm{UL}\star}+Kb_l^{\mathrm{DL}\star}$ is the number of tokens used by AP $l$ in the label, and minimize the composite loss
\begin{equation}
    \mathcal L=\lambda_{\rm t}\,\mathrm{KL}\!\big(\mathbf T^\star\,\|\,\mathbf T\big)
    +\lambda_{\rm r}\,\mathrm{BCE}_{a^\star}\!\big(q,q^\star\big)
    +\lambda_{\rm a}\,\mathrm{BCE}\!\big(p,a^\star\big),
\end{equation}
where the first term is the Kullback--Leibler divergence between the target and
predicted token distributions over the APs, the second is a binary
cross-entropy on the downlink ratio masked to the APs that are active in the
label, and the third is a binary cross-entropy on the activity head. Decoupling
the allocation into a distribution over APs, a per-AP uplink/downlink split, and
an on/off decision matches the structure of the greedy solution and guarantees that the budget constraint is satisfied by construction, while the on/off decision is learned explicitly rather than inferred from small predicted bit values. 
The network is trained with
the Adam optimizer using minibatches and early stopping on a validation set.
This parameterization follows the permutation-equivariant, size-generalizable
GNN design that has proven effective for radio resource
allocation~\cite{shen2021gnn,eisen2020optimal}.

\section{Spectral Efficiency With the Actual Quantizers}
\label{section:Bussgang}
The AQNM is an approximation of the true quantizer behavior. After
training, the learned bit-allocation policy is evaluated end-to-end under
spatially correlated Rayleigh fading with the actual Lloyd--Max quantizers applied to the uplink despread pilot signals and the downlink precoding vectors, instead of the AQNM used during optimization and training. This captures the actual quantization distortion, including its dependence on the quantizer input and its correlation across the antennas, thereby enabling a more reliable assessment of the spectral efficiency achieved by the learned bit-allocations.

\subsection{Uplink Pilot Quantization and Channel Estimation}

Each AP quantizes its despread pilot signals with the elementwise $b_l^{\rm UL}$-bit Lloyd--Max quantizer $\mathcal Q_l^{\rm UL}(\cdot)$, which is applied separately to the real and imaginary parts of each entry with a scale matched to its variance.
The CPU computes the channel estimates by applying the same AP-wise estimator as in Section~\ref{section:sysmod} to the actual quantized despread pilot signals $\widetilde{\mathbf y}_{t,l}=\mathcal Q_l^{\rm UL}(\mathbf y_{t,l})$, i.e., $\widehat{\mathbf h}_{kl}=\sqrt{\eta_k\tau_p}\,\alpha_{l}^{\rm UL}\mathbf R_{kl}(\mathbf \Psi_{t_k,l}^{\rm UL})^{-1}\widetilde{\mathbf y}_{t_k,l}$, and $\widehat{\mathbf h}_{k}=[\widehat{\mathbf h}_{k1}^{\Ttran}\ \cdots\ \widehat{\mathbf h}_{kL}^{\Ttran}]^{\Ttran}$.

\subsection{Exact Downlink Precoder Quantization}
Using these channel estimates
$\{\widehat{\mathbf h}_{k}\}_{k=1}^K$, the CPU computes centralized
precoding vectors $\mathbf w_k$, $k=1,\ldots,K$. For the downlink, the exact $b_l^{\rm DL}$-bit scalar quantizer, with the same per-antenna scale as in Section~\ref{section:sysmod}, is applied elementwise to the real and imaginary parts of $\mathbf W_l$, which gives the quantized precoding vectors
\begin{equation}
    \widetilde{\mathbf w}_{kl}
    =
    \mathcal Q_l^{\rm DL}(\mathbf w_{kl}),
    \qquad k=1,\ldots,K,
\end{equation}
and $\widetilde{\mathbf w}_k=[\widetilde{\mathbf w}_{k1}^{\Ttran}\ \cdots\ \widetilde{\mathbf w}_{kL}^{\Ttran}]^{\Ttran}$. Since AP $l$ knows $\widetilde{\mathbf W}_l$, it scales $\widetilde{\mathbf W}_l$ by a scalar, if needed, so that $\sum_{k}\|\widetilde{\mathbf w}_{kl}\|^2\leq P_{\max}$. In contrast to the quantization of precoded samples, the precoder quantization error remains fixed over the $\tau_d$ downlink channel uses and does not create a separate additive distortion term in the transmitted signal. Its effect is therefore captured exactly by using the quantized precoders directly in the spectral efficiency evaluation, and no linearization of the downlink quantizer is needed. The transmitted signal is
\begin{equation}
    \widetilde{\mathbf x}
    =
    \sum_{i=1}^{K}
    \widetilde{\mathbf w}_i s_i,
\end{equation}
and the received signal at UE $k$ is
\begin{align}
    y_k
    =
    \mathbf h_k^{\Htran}
    \widetilde{\mathbf w}_k s_k
    +
    \sum_{i\neq k}
    \mathbf h_k^{\Htran}
    \widetilde{\mathbf w}_i s_i
    +
    n_k.
\end{align}

\subsection{Ergodic Spectral Efficiency}
\label{sec:uatf_se}
Following the use-and-then-forget bounding technique, UE $k$ only uses the
average effective channel
\begin{equation}
    \mu_k
    =
    \mathbb E
    \left\{
        \mathbf h_k^{\Htran}
        {\widetilde{\mathbf w}_k}
    \right\}
\end{equation}
for coherent detection, while all remaining randomness is treated as
uncorrelated effective noise.
The received signal can therefore be rewritten as
\begin{align}
    y_k
    &=
    \mu_k s_k
    +
    \left(
        \mathbf h_k^{\Htran}
        {\widetilde{\mathbf w}_k}
        -
        \mu_k
    \right)s_k
    \nonumber\\
    &
    +
    \sum_{i\neq k}
    \mathbf h_k^{\Htran}
    {\widetilde{\mathbf w}_i} s_i
    +
    n_k.
\end{align}
The resulting effective SINR is given in \eqref{eq:uatf_sinr}, shown
at the top of the next page.
\begin{figure*}[!t]
\centering
\begingroup
\begin{equation}
\gamma_{k,{\rm UatF}}
=
\frac{
    |\mu_k|^2
}{
    \underbrace{
    \mathbb E
    \left\{
        \left|
        \mathbf h_k^{\Htran}
        {\widetilde{\mathbf w}_k}
        \right|^2
    \right\}
    -
    |\mu_k|^2
    }_{\text{beamforming uncertainty}}
    +
    \underbrace{
    \sum_{i\neq k}
    \mathbb E
    \left\{
        \left|
        \mathbf h_k^{\Htran}
        {\widetilde{\mathbf w}_i}
        \right|^2
    \right\}
    }_{\text{multi-user interference}}
    +
    \sigma^2
}.
\label{eq:uatf_sinr}
\end{equation}
\endgroup
\vspace{-0.5em}
\noindent\rule{\textwidth}{0.4pt}
\vspace{-1.0em}
\end{figure*}
Hence, an achievable ergodic spectral efficiency of UE $k$ is lower bounded by
\begin{equation}
    R_{k,{\rm UatF}}
    =
    \frac{\tau_d}{\tau_c}
    \log_2
    \left(
        1+\gamma_{k,{\rm UatF}}
    \right).
\end{equation}
The expectations in \eqref{eq:uatf_sinr} are computed via Monte Carlo
averaging over the channel realizations, where in each realization the exact
quantizers are applied to both the uplink despread pilot signals and the
downlink precoding vectors. This provides an end-to-end validation of the
learned bit-allocation policy that does not rely on the AQNM used during
label generation and training.
\section{Simulation Results}
\label{section:results}
In this section, we evaluate the performance of the proposed GNN-based joint uplink--downlink fronthaul bit-allocation scheme. 
We compare the proposed GNN-based scheme with the following
three bit-allocation methods:

\begin{itemize}
    \item \emph{Greedy:} The bit-allocation is obtained via the locally optimized allocation presented in Algorithm~\ref{alg:label_generation_compact}.

    \item \emph{Uniform:} The total weighted fronthaul budget is
    divided equally among the uplink and downlink allocation
    variables, independently of the channel conditions. The resulting
    continuous allocation is projected onto a feasible integer
    allocation satisfying the bit limits, total budget, and
    AP-consistency constraint.

    \item \emph{Heuristic:} The fronthaul budget is distributed
    according to the aggregate large-scale channel strength of each
    AP. The uplink and downlink importance scores are calculated
    according to \eqref{eq:aggregate_strengths}, converted into continuous budget fractions
    using \eqref{eq:token_budget_fraction}, and finally mapped to continuous bit resolutions
    according to \eqref{eq:initial_bit_allocation}. These resolutions are then projected onto a
    feasible integer allocation satisfying the bit limits, total
    budget, and AP-consistency constraint.
\end{itemize}
 
 We first compare the distributions of the sum spectral efficiency achieved by the four schemes under three different total fronthaul budgets. We then investigate the size generalization capability of the proposed GNN.
Finally, we compare the online bit-allocation runtimes of the four
schemes.
\subsection{Simulation Setup}

Unless otherwise stated, we consider a cell-free massive MIMO
network with $L=36$ APs, each equipped with $N=4$ antennas,
jointly serving $K=16$ single-antenna UEs. The APs are deployed on
a fixed regular grid over a $1\,\mathrm{km}\times1\,\mathrm{km}$
square area, while the UE locations are independently and uniformly
generated over the same area for every network realization. The AP
height relative to the UEs is $10$\,m. The carrier frequency is
$3.5$\,GHz, the system bandwidth is $50$\,MHz, and a noise figure of
$5$\,dB is used. The large-scale fading coefficient is modeled in dB as $10\log_{10}(\beta_{kl})=-32.4-20\log_{10}(f_c/1\,\mathrm{GHz})-31.9\log_{10}(d_{kl}/1\,\mathrm{m})+F_{kl}$~\cite{3GPP38901}, where $f_c$ is the carrier frequency, $d_{kl}$ is the three-dimensional distance between AP $l$ and UE $k$, $F_{kl}$ is the shadow fading. Log-normal shadow fading with a standard deviation
of $4$\,dB is independently generated for every AP--UE link.

Spatially correlated Rayleigh fading is considered using the local
scattering model \cite[Chapter~2]{massivemimobook}. Each AP employs a uniform linear array with
half-wavelength antenna spacing, and the azimuth and elevation angular
standard deviations are both set to $15^\circ$. The coherence block
contains $\tau_c=200$ channel uses, of which $\tau_p=8$ are allocated
to uplink pilot transmission. The pilots are assigned to the UEs in a
round-robin manner; hence, pilot reuse is employed when $K>\tau_p$.
The pilot transmit power is $\eta_k=0.1$\,W for every
UE, and the maximum downlink transmit power is
$P_{\max}=2$\,W for every AP. Centralized regularized zero-forcing
precoding is employed with regularization parameter $10^{-2}$. All precoding vectors are scaled by a common factor such that the most loaded active AP satisfies its power constraint with equality under the AQNM, i.e., $\max_l \alpha_l^{\rm DL}\mathrm{tr}(\mathbf C_{x,l})=P_{\max}$. In the evaluation with the actual quantizers, an AP whose quantized precoders exceed $P_{\max}$ additionally scales them down.

The fronthaul quantization resolution is restricted to
$b\in\{0,1,\ldots,6\}$ bits, i.e., $b_{\max}=6$, where $b=0$ corresponds to deactivating
the associated AP. The same uplink resolution is used for all pilot
signals quantized at a given AP. Therefore, one additional uplink bit
at an AP consumes $\tau_p$ budget units, whereas one additional
downlink bit consumes $K$ budget units. All considered allocation schemes satisfy the AP-consistency
constraint introduced in Section~\ref{sec:greedy_label_generation}.

The supervised GNN is trained using $5000$ independently generated
large-scale fading realizations, while another $500$ realizations are
used for validation. The greedy algorithm is used to generate the
training labels. For each considered allocation, its AQNM-based
sample-average objective is evaluated over $50$ common small-scale
fading realizations. The GNN contains five message-passing layers with
a hidden dimension of $128$. It is trained using the Adam optimizer
with an initial learning rate of $2\times10^{-3}$, which is halved every $250$ epochs, a weight decay of $10^{-5}$, a minibatch size of
$1000$, and a maximum of $1000$ epochs. The loss weights are $\lambda_{\rm t}=L$ and $\lambda_{\rm r}=\lambda_{\rm a}=1$. Validation is performed every
five epochs, and training is terminated early if the validation loss
does not improve for ten consecutive validation checks.

The final performance is evaluated over $20$ independently generated test setups. For each setup, the proposed GNN scheme is compared with the greedy, uniform, and channel-strength-based heuristic allocation schemes described earlier. The final spectral efficiencies are evaluated with the actual Lloyd--Max quantizers as described in Section~\ref{section:Bussgang}. The use-and-then-forget
spectral efficiency is estimated using $100$ independent small-scale
fading realizations. Thus, the results for each allocation scheme and system
configuration are based on $20\times100=2000$ independently
generated channel realizations. 
\subsection{Numerical Results}
\begin{figure*}[t]
    \centering

    \begin{subfigure}[t]{0.32\textwidth}
        \centering
        \includegraphics[width=\linewidth]{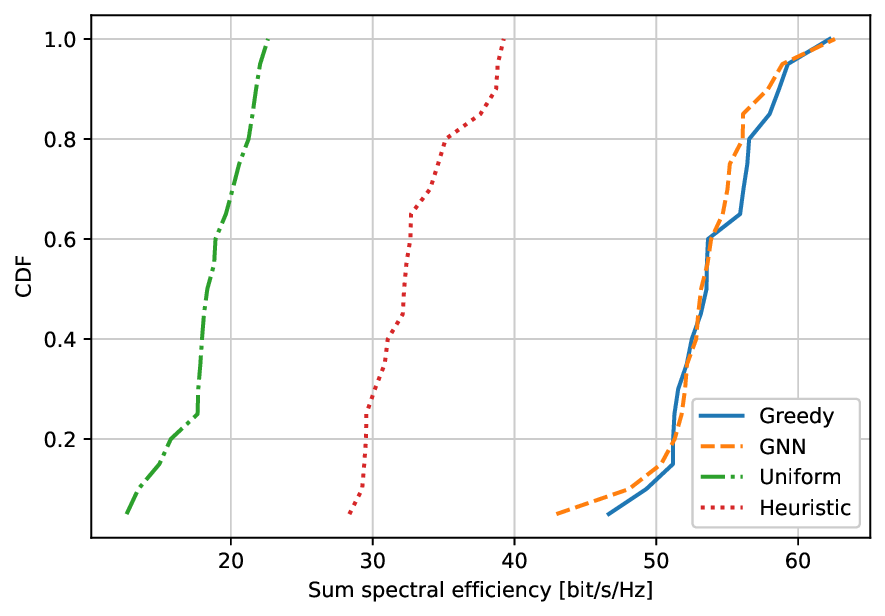}
    \caption{{$B_{\mathrm{tot}}=9$}}
        \label{fig:cdf_b180}
    \end{subfigure}
    \hfill
    \begin{subfigure}[t]{0.32\textwidth}
        \centering
        \includegraphics[width=\linewidth]{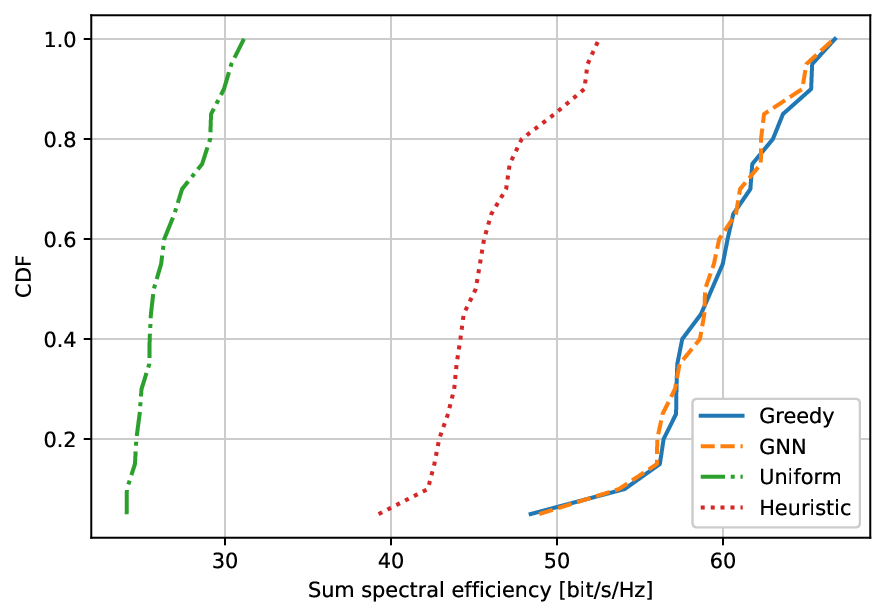}
    \caption{{$B_{\mathrm{tot}}=12$}}
        \label{fig:cdf_b240}
    \end{subfigure}
    \hfill
    \begin{subfigure}[t]{0.32\textwidth}
        \centering
        \includegraphics[width=\linewidth]{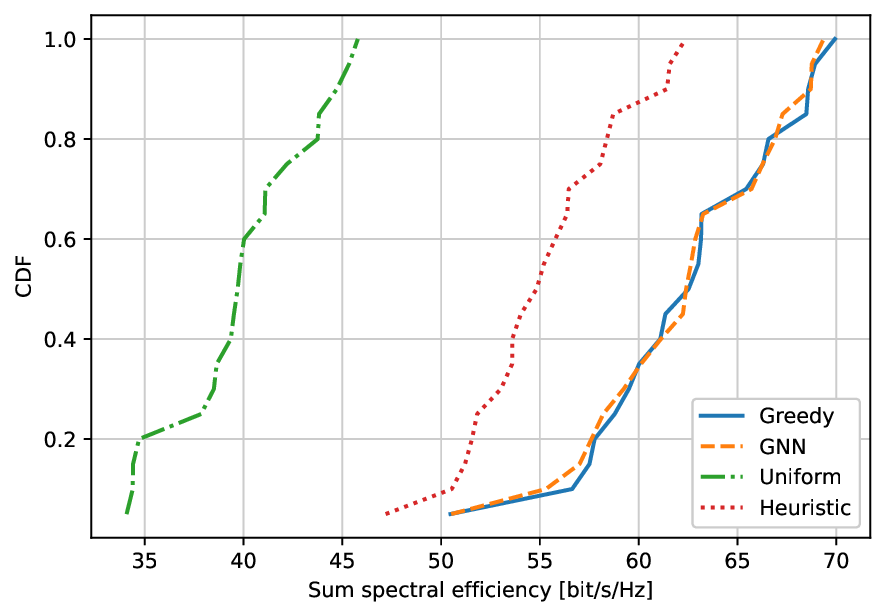}
    \caption{{$B_{\mathrm{tot}}=15$}}
        \label{fig:cdf_b300}
    \end{subfigure}

    \caption{CDF of the sum spectral efficiency for different total
    fronthaul bit budgets.}
    \label{fig:cdf_different_budgets}
\end{figure*}

Fig.~\ref{fig:cdf_different_budgets} compares the cumulative distribution functions (CDFs) of the
sum spectral efficiency for the considered allocation schemes under
different total fronthaul budgets. We can see that the uniform scheme exhibits the
lowest performance in all cases, 
because it distributes the available weighted budget independently
of the channel conditions and therefore cannot prioritize the APs
that contribute most strongly to channel estimation and downlink
transmission. The channel-strength-based heuristic performs better
because it assigns a larger fraction of the budget to APs having
stronger aggregate large-scale channels. Nevertheless, it does not
account for the joint effects of uplink quantization, pilot
contamination, multiuser interference, downlink precoding, and
fronthaul quantization. The proposed GNN-based scheme substantially
outperforms both baselines by learning these joint dependencies from
the locally optimized allocations generated by Algorithm~1. Note that Algorithm~\ref{alg:label_generation_compact}
provides a locally optimized reference allocation rather than a
globally optimal solution. Furthermore, it optimizes an AQNM-based
sample-average objective, whereas the reported results use the
actual Lloyd--Max quantizers. The GNN may therefore
occasionally outperform the greedy method, as observed in the figures.
\begin{figure}[t]
    \centering\includegraphics[width=0.95\columnwidth]
        {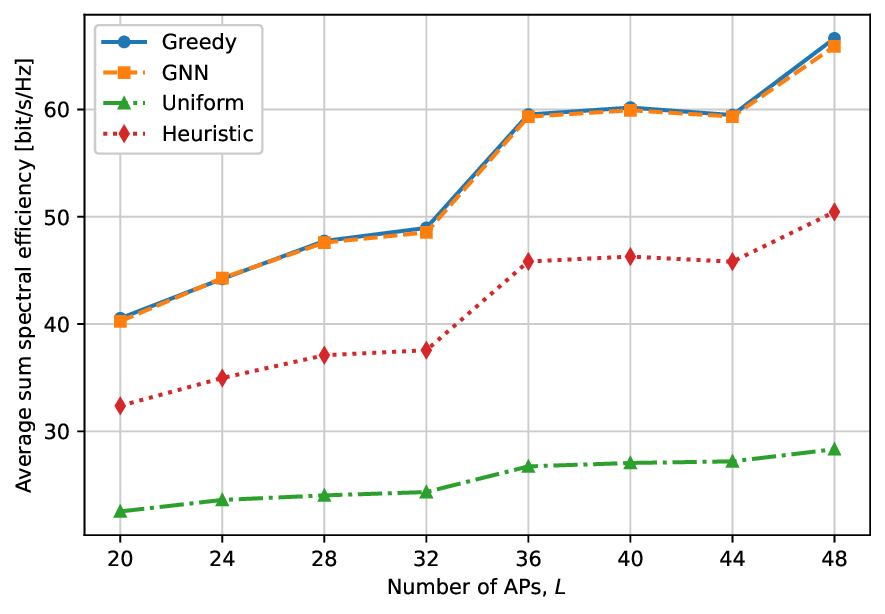}
    \caption{Average sum spectral efficiency versus the number of APs.
    The GNN is trained for $L=36$, $K=16$, and $B_{\mathrm{tot}}=12$,
    and evaluated at different numbers of APs without retraining.}
    \label{fig:generalization_L}
\end{figure}

Fig.~\ref{fig:generalization_L} investigates the generalization
capability of the proposed GNN with respect to the number of APs.
The GNN is trained only for
$(L,K,B_{\mathrm{tot}})=(36,16,12)$ and is subsequently evaluated,
without retraining, for
$L\in\{20,24,\ldots,48\}$. For each $L$, the APs are placed row by row on the first $L$ points of a regular grid with $\lceil\sqrt{L}\rceil$ columns, so that the last row is only partially occupied when $L$ is not a multiple of $\lceil\sqrt{L}\rceil$. To maintain approximately the same
fronthaul budget per AP as in the training configuration, the total
budget is scaled according to
\begin{equation}
    B_{\mathrm{tot}}(L)
    =
    \frac{12L}{36}.
    \label{eq:budget_scaling_L}
\end{equation}
Similar to what we have observed in Fig.~\ref{fig:cdf_different_budgets}, the uniform allocation
provides the lowest sum spectral efficiency because it does not adapt
to the channel conditions, whereas the channel-strength-based
heuristic achieves better performance by prioritizing APs with
stronger aggregate large-scale channels. The proposed GNN consistently
outperforms these two baselines and follows the performance trend of
the greedy reference method over the entire considered range of $L$.
 This demonstrates that
the shared message-passing operations and permutation-equivariant
architecture enable the GNN trained at $L=36$ to operate effectively
on networks with both fewer and more APs without retraining.

\begin{figure}[t]
    \centering\includegraphics[width=0.95\columnwidth]{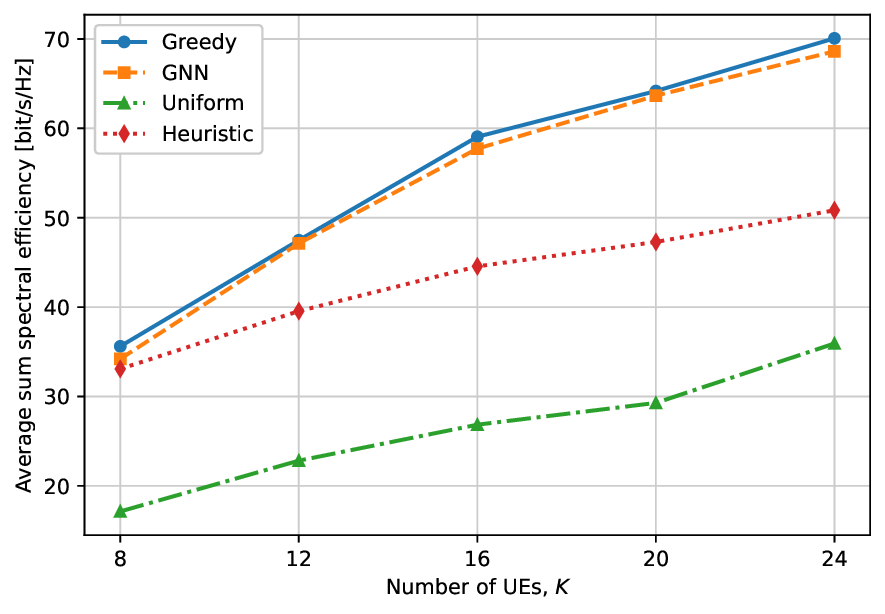}
    \caption{Average sum spectral efficiency versus the number of UEs.
    The GNN is trained for $L=36$, $K=16$, and
    $B_{\mathrm{tot}}=12$, and evaluated at different numbers of UEs
    without retraining. }
    \label{fig:generalization_K}
\end{figure}

Fig.~\ref{fig:generalization_K} evaluates the generalization
capability of the proposed GNN with the number of UEs.
The GNN is trained for
$(L,K,B_{\mathrm{tot}})=(36,16,12)$ and is evaluated for $K\in\{8,12,16,20,24\}$ without retraining, while keeping $L=36$ fixed.
To keep the same total fronthaul budget per UE as in the training
configuration, the budget is scaled proportionally with $K$ as
\begin{equation}
    B_{\mathrm{tot}}(K)
    =
    \frac{12K}{16}.
    \label{eq:budget_scaling_K}
\end{equation}
The GNN follows the overall performance trend of the greedy reference
method and outperforms the uniform and
channel-strength-based heuristic schemes over the entire considered
range of $K$, demonstrating its ability to operate with a number of
UEs different from that used during training. Overall, the GNN maintains a clear advantage over the two
baseline schemes without requiring retraining.

\begin{figure}[t]
    \centering\includegraphics[width=0.95\columnwidth]{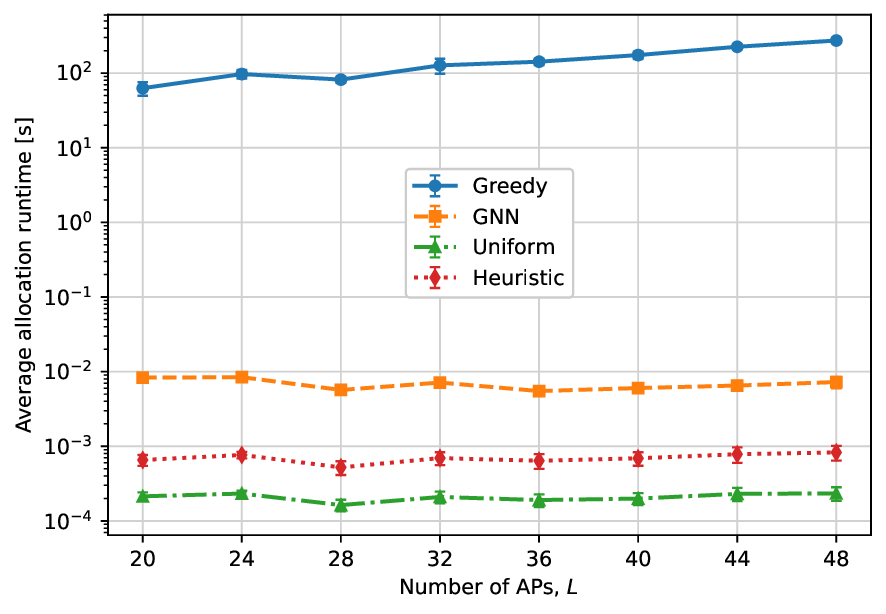}
    \caption{Average online bit-allocation runtime versus the number of APs.}
    \label{fig:runtime_vs_L}
\end{figure}

In Fig.~\ref{fig:runtime_vs_L}, we evaluate  the average online
bit-allocation runtimes of the considered schemes as the number of
APs increases. Similar to the generalization experiment in
Fig.~\ref{fig:generalization_L}, the total fronthaul budget is scaled
with $L$ according to \eqref{eq:budget_scaling_L}, with
$B_{\mathrm{tot}}=12$ at the reference configuration $L=36$. The
reported GNN runtime includes feature construction, neural network
inference, and projection of the continuous network outputs onto a
feasible integer allocation, while excluding offline training and
model loading. The GNN requires substantially less online computation
than the greedy method, with an average runtime of less than
$10$\,ms over the considered values of $L$. Moreover, its runtime
remains relatively stable as the network size increases. In contrast,
the greedy method requires approximately $80$--$300$\,s and generally
becomes more computationally demanding as $L$ increases because a
larger number of allocation variables and feasible exchange operations
must be examined. This substantial reduction in runtime is achieved
while the GNN retains sum spectral efficiency close to that of the
greedy method, as observed in Fig.~\ref{fig:generalization_L}. The
uniform and channel-strength-based heuristic schemes have lower
runtimes than the GNN-based allocation because they do not perform an iterative search or
neural network inference; however, the previous results have shown that
this computational advantage is accompanied by considerably lower
spectral efficiency.

\section{Conclusion}
\label{section:conclusion}

This paper investigated joint uplink--downlink fronthaul bit-allocation for
centralized cell-free massive MIMO and developed a GNN-based approach for
efficiently approximating locally optimized allocations. A greedy algorithm
was used to generate reference solutions, while the proposed bipartite GNN
learned to infer the corresponding allocations directly from large-scale
fading information.

The numerical results showed that the GNN-based allocation and 
the locally optimized greedy approach achieve very similar spectral
efficiency at all the considered setups. At the same time, the GNN requires
orders of magnitude less online computation than the greedy algorithm. The
results also demonstrated that a GNN trained for a particular number of APs
and UEs can generalize effectively to moderately different network sizes
without retraining. Moreover, the GNN-based solution achieves significantly higher spectral efficiency than the uniform and heuristic schemes.

\bibliographystyle{IEEEtran}
\bibliography{IEEEabrv,refs}

\begin{thebibliography}{10}
\providecommand{\url}[1]{#1}
\csname url@samestyle\endcsname
\providecommand{\newblock}{\relax}
\providecommand{\bibinfo}[2]{#2}
\providecommand{\BIBentrySTDinterwordspacing}{\spaceskip=0pt\relax}
\providecommand{\BIBentryALTinterwordstretchfactor}{4}
\providecommand{\BIBentryALTinterwordspacing}{\spaceskip=\fontdimen2\font plus
\BIBentryALTinterwordstretchfactor\fontdimen3\font minus
  \fontdimen4\font\relax}
\providecommand{\BIBforeignlanguage}[2]{{%
\expandafter\ifx\csname l@#1\endcsname\relax
\typeout{** WARNING: IEEEtran.bst: No hyphenation pattern has been}%
\typeout{** loaded for the language `#1'. Using the pattern for}%
\typeout{** the default language instead.}%
\else
\language=\csname l@#1\endcsname
\fi
#2}}
\providecommand{\BIBdecl}{\relax}
\BIBdecl

\bibitem{Ngo2017b}
H.~Q. Ngo, A.~Ashikhmin, H.~Yang, E.~G. Larsson, and T.~L. Marzetta,
  ``Cell-free {Massive} {MIMO} versus small cells,'' \emph{{IEEE} Trans.
  Wireless Commun.}, vol.~16, no.~3, pp. 1834--1850, 2017.

\bibitem{Nayebi2017a}
E.~Nayebi, A.~Ashikhmin, T.~L. Marzetta, H.~Yang, and B.~D. Rao, ``Precoding
  and power optimization in cell-free {Massive} {MIMO} systems,'' \emph{{IEEE}
  Trans. Wireless Commun.}, vol.~16, no.~7, pp. 4445--4459, 2017.

\bibitem{Interdonato2018}
G.~Interdonato, E.~Bj\"ornson, H.~Q. Ngo, P.~Frenger, and E.~G. Larsson,
  ``Ubiquitous cell-free massive {MIMO} communications,'' \emph{{EURASIP} J.
  Wirel. Commun. Netw.}, vol. 2019, no. 197, 2019.

\bibitem{ozdogan2019performance}
O.~\"Ozdogan, E.~Bj\"ornson, and J.~Zhang, ``Performance of cell-free massive
  {MIMO} with {Rician} fading and phase shifts,'' \emph{{IEEE} Trans. Wireless
  Commun.}, vol.~18, no.~11, pp. 5299--5315, 2019.

\bibitem{Bjornson2020a}
E.~Björnson and L.~Sanguinetti, ``Scalable cell-free massive {MIMO} systems,''
  \emph{IEEE Trans. Commun.}, vol.~68, no.~7, pp. 4247--4261, 2020.

\bibitem{Park2013joint}
S.-H. Park, O.~Simeone, O.~Sahin, and S.~Shamai, ``Joint precoding and
  multivariate backhaul compression for the downlink of cloud radio access
  networks,'' \emph{IEEE Trans. Signal Process.}, vol.~61, no.~22, pp.
  5646--5658, 2013.

\bibitem{Zhou2014optimized}
Y.~Zhou and W.~Yu, ``Optimized backhaul compression for uplink cloud radio
  access network,'' \emph{IEEE J. Sel. Areas Commun.}, vol.~32, no.~6, pp.
  1295--1307, 2014.

\bibitem{Bashar2019b}
M.~{Bashar}, K.~{Cumanan}, A.~G. {Burr}, H.~Q. {Ngo}, M.~{Debbah}, and
  P.~{Xiao}, ``Max–min rate of cell-free {Massive} {MIMO} uplink with optimal
  uniform quantization,'' \emph{{IEEE} Trans. Commun.}, vol.~67, no.~10, pp.
  6796--6815, 2019.

\bibitem{Masoumi2020a}
H.~{Masoumi} and M.~J. {Emadi}, ``Performance analysis of cell-free {Massive}
  {MIMO} system with limited fronthaul capacity and hardware impairments,''
  \emph{{IEEE} Trans. Wireless Commun.}, vol.~19, no.~2, pp. 1038--1053, 2020.

\bibitem{Boroujerdi2019a}
M.~N. Boroujerdi, A.~Abbasfar, and M.~Ghanbari, ``Cell free massive {MIMO} with
  limited capacity fronthaul,'' \emph{Wireless Pers. Commun.}, vol. 104, pp.
  633--648, 2019.

\bibitem{Topal2024energy}
O.~A. Topal, {\"O}.~T. Demir, E.~Bj{\"o}rnson, and C.~Cavdar,
  ``Energy-efficient cell-free massive {MIMO} with wireless fronthaul,'' in
  \emph{Proc. Asilomar Conf. Signals, Syst., Comput.}, 2024, pp. 1591--1596.

\bibitem{demir2026joint}
{\"O}.~T. Demir, ``Joint quantized precoding and bit allocation for
  fronthaul-constrained cell-free massive {MIMO},'' \emph{arXiv preprint
  arXiv:2609.26149}, 2026.

\bibitem{khorsandmanesh2023optimized}
Y.~Khorsandmanesh, E.~Bj{\"o}rnson, and J.~Jald{\'e}n, ``Optimized precoding
  for {MU-MIMO} with fronthaul quantization,'' \emph{IEEE Trans. Wireless
  Commun.}, vol.~22, no.~11, pp. 7102--7115, 2023.

\bibitem{Ramezani2024MSE}
P.~Ramezani, Y.~Khorsandmanesh, and E.~Bj\"ornson, ``{MSE} minimization in
  {RIS}-aided {MU-MIMO} with discrete phase shifts and fronthaul
  quantization,'' in \emph{Proc. IEEE Veh. Technol. Conf. (VTC-Spring)}, 2024,
  pp. 1--5.

\bibitem{Ramezani2025joint}
------, ``Joint discrete precoding and {RIS} optimization for {RIS}-assisted
  {MU-MIMO} communication systems,'' \emph{IEEE Trans. Commun.}, vol.~73,
  no.~3, pp. 1531--1546, 2025.

\bibitem{Khorsandmanesh2026splitting}
Y.~Khorsandmanesh, E.~Bj\"ornson, and J.~Jald{\'e}n, ``Splitting precoding with
  subspace selection and quantized refinement for massive {MIMO},'' \emph{IEEE
  Wireless Commun. Lett.}, vol.~15, pp. 3129--3133, 2026.

\bibitem{demir2026GNN}
{\"O}.~T. {Demir} and P.~{Ramezani}, ``{GNN}-based joint split processing and
  fronthaul bit allocation for cell-free massive {MIMO},'' \emph{IEEE Wireless
  Commun. Lett.}, 2026, submitted.

\bibitem{Lee2018resource}
W.~Lee, ``Resource allocation for multi-channel underlay cognitive radio
  network based on deep neural network,'' \emph{IEEE Commun. Lett.}, vol.~22,
  no.~9, pp. 1942--1945, 2018.

\bibitem{Sun2018learning}
H.~Sun, X.~Chen, Q.~Shi, M.~Hong, X.~Fu, and N.~D. Sidiropoulos, ``Learning to
  optimize: Training deep neural networks for interference management,''
  \emph{IEEE Trans. Signal Process.}, vol.~66, no.~20, pp. 5438--5453, 2018.

\bibitem{eisen2020optimal}
M.~Eisen and A.~Ribeiro, ``Optimal wireless resource allocation with
  application to random edge graph neural networks,'' \emph{IEEE Trans. Signal
  Process.}, vol.~68, pp. 2977--2991, 2020.

\bibitem{shen2021gnn}
Y.~Shen, Y.~Shi, J.~Zhang, and K.~B. Letaief, ``Graph neural networks for
  scalable radio resource management: Architecture design and theoretical
  analysis,'' \emph{IEEE J. Sel. Areas Commun.}, vol.~39, no.~1, pp. 101--115,
  2021.

\bibitem{Lee2021graph}
M.~Lee, G.~Yu, and G.~Y. Li, ``Graph embedding-based wireless link scheduling
  with few training samples,'' \emph{IEEE Trans. Wireless Commun.}, vol.~20,
  no.~4, pp. 2282--2294, 2021.

\bibitem{Orhan2015a}
O.~Orhan, E.~Erkip, and S.~Rangan, ``Low power analog-to-digital conversion in
  millimeter wave systems: Impact of resolution and bandwidth on performance,''
  in \emph{IEEE ITA}, 2015.

\bibitem{3GPP38901}
{3GPP}, ``Study on channel model for frequencies from 0.5 to 100 {GHz},'' 3rd
  Generation Partnership Project (3GPP), Technical Report TR 38.901, 2020,
  version 16.1.0.

\bibitem{massivemimobook}
E.~Bj\"{o}rnson, J.~Hoydis, and L.~Sanguinetti, ``Massive {MIMO} networks:
  {Spectral}, energy, and hardware efficiency,'' \emph{Foundations and
  Trends{\textregistered} in Signal Processing}, vol.~11, no. 3-4, pp.
  154--655, 2017.

\end{thebibliography}
\end{document}